\documentclass[sigconf]{acmart}

\usepackage{graphicx}
\usepackage{soul}
\usepackage[skip=0pt]{caption}
\usepackage{subcaption}
\usepackage{rotating}
\usepackage{xspace}
\usepackage{color, colortbl}
\usepackage{xcolor}
\usepackage{minted}
\usepackage{multirow}
\usepackage[many]{tcolorbox}
\usepackage{enumitem}
\usepackage{ulem}

\newminted[mintedjava]{java}{
  frame=lines,
  framesep=1mm,
  baselinestretch=1,
  fontsize=\scriptsize,
  breaklines,
  breakanywhere,
  linenos,
  numbersep=4pt,
  style=xcode
}
\newcounter{findingctr}
\newtcolorbox{finding}{%
  colback=blue!5,        
  colframe=blue!50!black,
  boxrule=0.4pt,         
  arc=0.8mm,             
left=0pt, right = 0pt, top=0pt, bottom=0pt,
  before upper={\stepcounter{findingctr}Finding~\thefindingctr:~}, 
}

\AtBeginDocument{%
  }

\setcopyright{none}
\renewcommand\footnotetextcopyrightpermission[1]{}

\acmConference[2026]{}{}{}

\usepackage{rotating}
\usepackage{array}
\usepackage{booktabs}
\usepackage{ragged2e}
\usepackage{afterpage}
\usepackage{tabularx} 
\usepackage{array}
\usepackage{makecell}
\usepackage{afterpage}
\usepackage{balance}

\newcolumntype{L}[1]{>{\RaggedRight\arraybackslash}p{#1}}
\newcolumntype{C}[1]{>{\Centering\arraybackslash}p{#1}}
\definecolor{claudeorange}{rgb}{0.8,0.48,0.35}
\begin{document}

\title{Usage, Effects and Requirements for AI Coding Assistants in the Enterprise: An Empirical Study}

\author{Maja Vukovic}
\authornotemark[1]
\affiliation{%
  \institution{IBM Research}
  \city{Yorktown Heights}
  \state{NY}
  \country{USA}
}
\email{maja@us.ibm.com}

\author{Rangeet Pan}
\affiliation{%
  \institution{IBM Research}
  \city{Yorktown Heights}
  \state{NY}
  \country{USA}
  }
\email{rangeet.pan@ibm.com}

\author{Tin Kam Ho}
\affiliation{%
  \institution{IBM Research}
  \city{Yorktown Heights}
  \state{NY}
  \country{USA}
  }
\email{tho@us.ibm.com}

\author{Rahul Krishna}
\affiliation{%
  \institution{IBM Research}
  \city{Yorktown Heights}
  \state{NY}
  \country{USA}
  }
\email{rkrsn@ibm.com}

\author{Raju Pavuluri}
\affiliation{%
  \institution{IBM Research}
  \city{Yorktown Heights}
  \state{NY}
  \country{USA}
  }
\email{pavuluri@us.ibm.com}

\author{Michele Merler}
\affiliation{%
  \institution{IBM Research}
  \city{Yorktown Heights}
  \state{NY}
  \country{USA}
  }
\email{mimerler@us.ibm.com}

\renewcommand{\shortauthors}{Vukovic et al.}

\begin{abstract}
The rise of large
language models (LLMs) has accelerated the development of automated techniques and tools for supporting various software engineering
tasks, e.g., program understanding, code generation, software testing, and program repair. As CodeLLMs are being employed
toward automating these tasks, one question that arises, especially in enterprise settings, is whether these coding assistants and the code LLMs
that power them are ready for real-world projects and enterprise use cases, and how do they impact the existing software engineering process and user experience. In this paper we survey 57 developers from different domains and with varying software engineering skill about their experience with AI coding assistants and CodeLLMs. We also reviewed 35 user surveys on the usage, experience
and expectations of professionals and students using AI coding assistants and CodeLLMs. Based on our study findings and analysis of existing surveys, we discuss the requirements for AI-powered
coding assistants.
\end{abstract}



\keywords{AI Coding Assistant, Survey, Usability Study, CodeLLM}


\maketitle
\section{Introduction}


 The evolution of modern software engineering has consistently leaned toward automation so as
to increase efficiency and reduce repetitive and low-level development tasks. Over the past decade,
various techniques ranging from the evolution of scripting utilities to more sophisticated development frameworks have attempted to simplify common software development tasks~\cite{Forsgren2021}. The advent of Code Large Language Models (CodeLLM) powered AI-coding assistants represents yet another
promising step in this direction, allowing developers to automate code generation, suggest com-
pletions, produce tests, and even modernize legacy codebases. The use of large language models
(LLMs) allows these AI assistants to enable contextually relevant guidance, streamline code
creation, and offer assistance in learning new programming languages. More recently, the rise of agentic systems has enabled further automation of more complex software engineering tasks such as bug fixing~\cite{he25agentsurvey}.



In this paper we survey 57 industry developers about their experience with AI coding assistants. The surveyed group came from several functional domains (e.g. software, consulting, research, etc.), with varying levels of software engineering experience. The intent of this survey was to understand the following:
\begin{itemize}[leftmargin=*]
\item What are the different utilization of coding assistants and desired use cases across the functional verticals
\item How does the development experience influence the expectations of an AI coding assistant. 
\item What are the differences in software engineering processes introduced by AI coding assistants.
\item What is the overall perceived effectiveness of the assistants
\item What are the requirements for the future AI coding assistants. 
\end{itemize}

Stemming from the user study, we look for requirements that will shape the research road map and features for future iterations of coding assistants. The contribution of our work is a survey that provides a comparative study of usage of AI coding assistants across a diverse set of users, understanding the impact of that usage in terms of organizational structure, productivity and quality.

\section{Related Work}

Research on AI coding assistants has grown rapidly, with GitHub Copilot, ChatGPT, Tabnine, and enterprise specific tools such as Amazon Q and IBM Watsonx Code Assistant among the most studied \cite{stack-overflow-survey-2024}. Large surveys \cite{weisz2025examininguseimpactai, ziegler2024measuring, sergeyuk2025using, stalnaker2025developer} and randomized controlled trials \cite{lyu2025myproductivityboosted, cui2024productivity, butler2024deardiaryrandomizedcontrolled, paradis2025icseseip} consistently show perceived productivity gains \cite{weber2024gain}, often in the range of 12--25\% speedups and up to almost one third of developers code being written with help from the AI tool \cite{liang2024large}. Industry case studies further validate these benefits in real-world settings \cite{bakal2025experiencegithubcopilotdeveloper, takerngsaksiri2025codereadabilityagelarge}. Major categories with satisfactory outcomes include code generation and refactoring, testing, code exaplanation and addressing data analytics issues \cite{das2024developersengagechatgptissuetracker}.

While productivity advantages are clear, the extent of improvement varies across tasks type and complexity \cite{wang2024rocks}, developer expertise \cite{gambacorta2024generative}, and organizational contexts. Less experienced and causal developers report higher adoption rates of AI coding tools and greater productivity gains \cite{cui2025effects}, using them to reduce opportunity costs. In open source, evidence has been found that this enhanced development activity translated even into increased short to medium-term career benefits for developers \cite{li2025assessing}. On the other hand recent studies \cite{becker2025measuringimpactearly2025ai} have found discrepancies between perceived and measured productivity for experienced developers.


Beyond efficiency, concerns about quality and security remain salient. Empirical audits show that Copilot-generated code often contains vulnerabilities, particularly in security-critical domains \cite{Perry_2023, klemmer2024using, fu2025security, tihanyi2025how}. Perceptions are similarly mixed: developers report faster delivery but raise doubts about maintainability, integration and correctness \cite{li2025vibecodinguxdesign,martinovic2025perceived,vaillant2024developersperceptionsimpactchatgpt}. Studies on readability and naming practices find no consistent degradation but highlight variability depending on context \cite{takerngsaksiri2025codereadabilityagelarge, lee2024predictability, zhou2025exploring}. Overall, the literature underscores that productivity gains may come at the cost of quality assurance and over-reliance is identified as a risk factor \cite{ahuchogu2025evaluating}.

Adoption and trust \cite{sabouri2025trust,omidvar2024evaluating} have emerged as central challenges. 
Legal and ethical questions compound these issues, as many practitioners remain unclear about copyright and licensing implications \cite{stalnaker2025developer,bahn2025copiloting}. 
Recent studies \cite{miller2025maybeneedexamplesindividual} found that widespread organizational expectations in industries for rapid productivity gains are limited by uneven adoption and familiarity with the tools from different team members, thus advocating for needed investment in learning.
Studies of learning and collaboration reveal similarly dual effects \cite{llerenaizquierdo2024towards}. Novices benefit from scaffolding and improved comprehension of large codebases \cite{shah2025students, prather2024widening, mailach2025ok,brown2025howzat}, yet risk shallow learning and over-reliance \cite{akhoroz2025conversationalaicodingassistant,ahuchogu2025evaluating}. Professionals likewise report efficiency gains but worry about deskilling in core competencies, even as new skills around prompting and verification emerge \cite{crowston2025deskilling, stray2025generative, vazpereira2025exploring}. 

The socio-economical consequences of the wide adoption of AI based coding tools are also being researched, with focus not only on workflow efficiency but also its potential for developer replacement \cite{kuhail2024replace} and displacement \cite{bahmed2024integrated}.
The use of AI coding assistants beyond software engineers has been largely understudied, with few exceptions discovering for example that scientists often use code generating models as an information retrieval tool for navigating unfamiliar programming languages and libraries \cite{obriedn2025scientists}.



Despite these advances, our analysis of prior studies reveals significant gaps. First, most of the studies focus on immediate productivity or single specific tasks \cite{nikolov2025googleusingaiinternal, corso2024generating2,mohamed2025impactllmassistantssoftwaredeveloper} rather than long-term maintainability, software quality, or organizational outcomes. Second, despite the recent proliferation of AI tools (Cursor, Windsurf, Replit, Zencoder, Cody, BlackboxAI, abacus AI, Qodo, Bolt etc.) studies have predominantly focused only on very few, namely ChatGPT and Github Copilot. 
Third, participants are often drawn from homogeneous cohorts, restricting generalizability across industries, regions, and developer demographics. 
These omissions highlight the need for more diverse, longitudinal, and governance-focused studies to inform both research and practice.

\section{Survey on AI coding assistants use in enterprise and academia}

\begin{figure}[t]
\centering
\includegraphics[width=\linewidth]{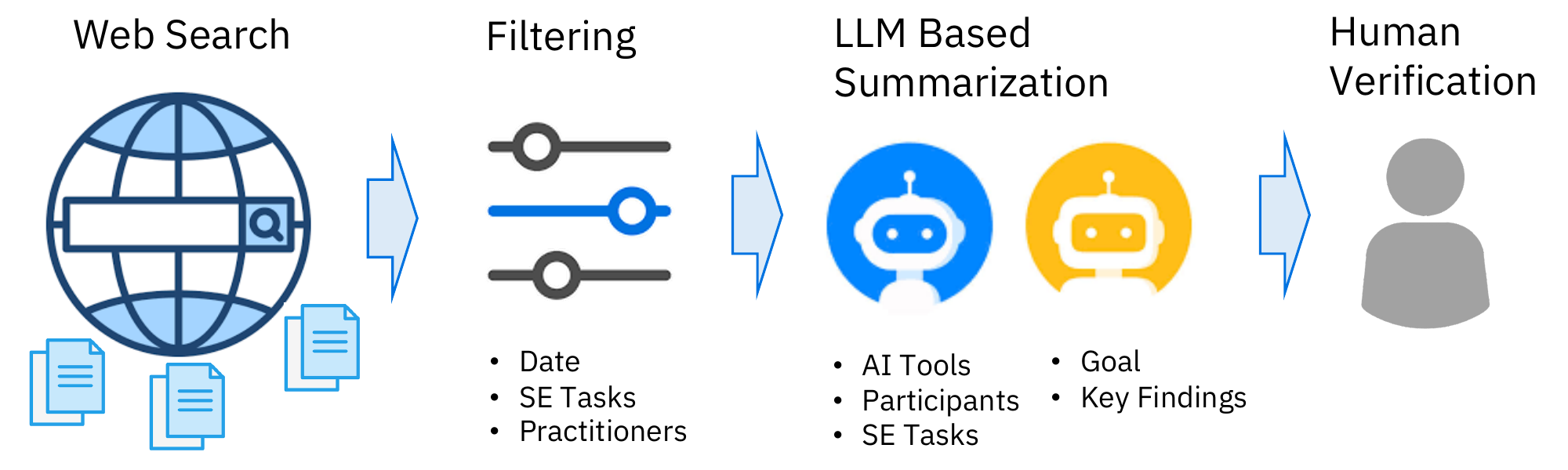}
\caption{Surveys search, selection and analysis process.}
\label{fig:survey_selection}
\end{figure}

In this Section we review 35 user surveys on the usage, experience and expectations of professionals and students using AI coding assistants and codeLLMs. While the focus of our work is on industry use-cases, we expanded the scope of the reviewed surveys to include academia as a comparison point. 
In the following we report the insights from this review, starting with the description of the methods used to search and collect the papers, the filtering criteria, the breakdown of AI tools and models investigated, the software engineering tasks involved, the participants information and goal of the surveys. Finally we report the collective key insights. A summary of the process is illustrated in Figure \ref{fig:survey_selection}.

\begin{figure*}[t]
\centering
\begin{subfigure}{0.45\textwidth}
\centering
\includegraphics[width=\linewidth]{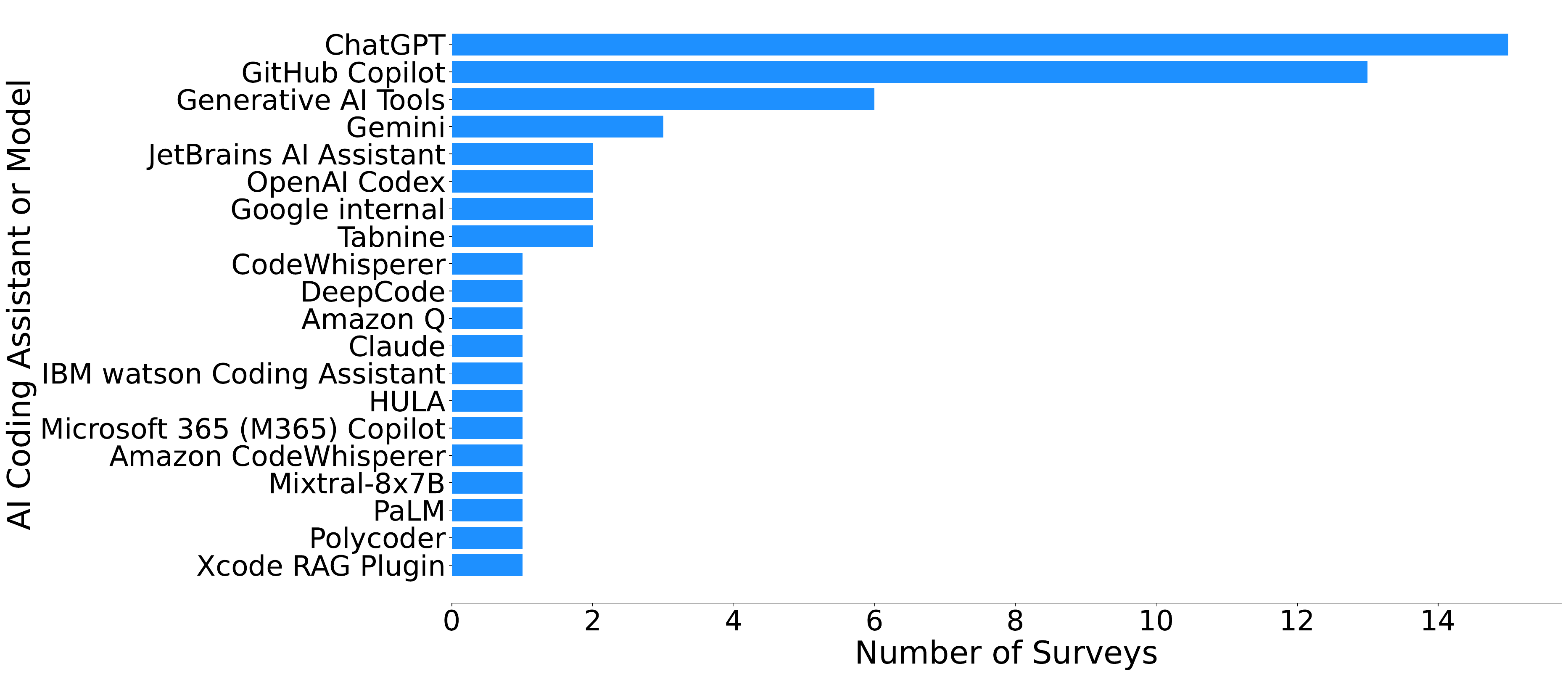}
\caption{AI Coding Assisting Tools}
\label{fig:comparison_ai_assistants_distribution}
\end{subfigure}
\begin{subfigure}{0.45\textwidth}
\centering
\includegraphics[width=\linewidth]{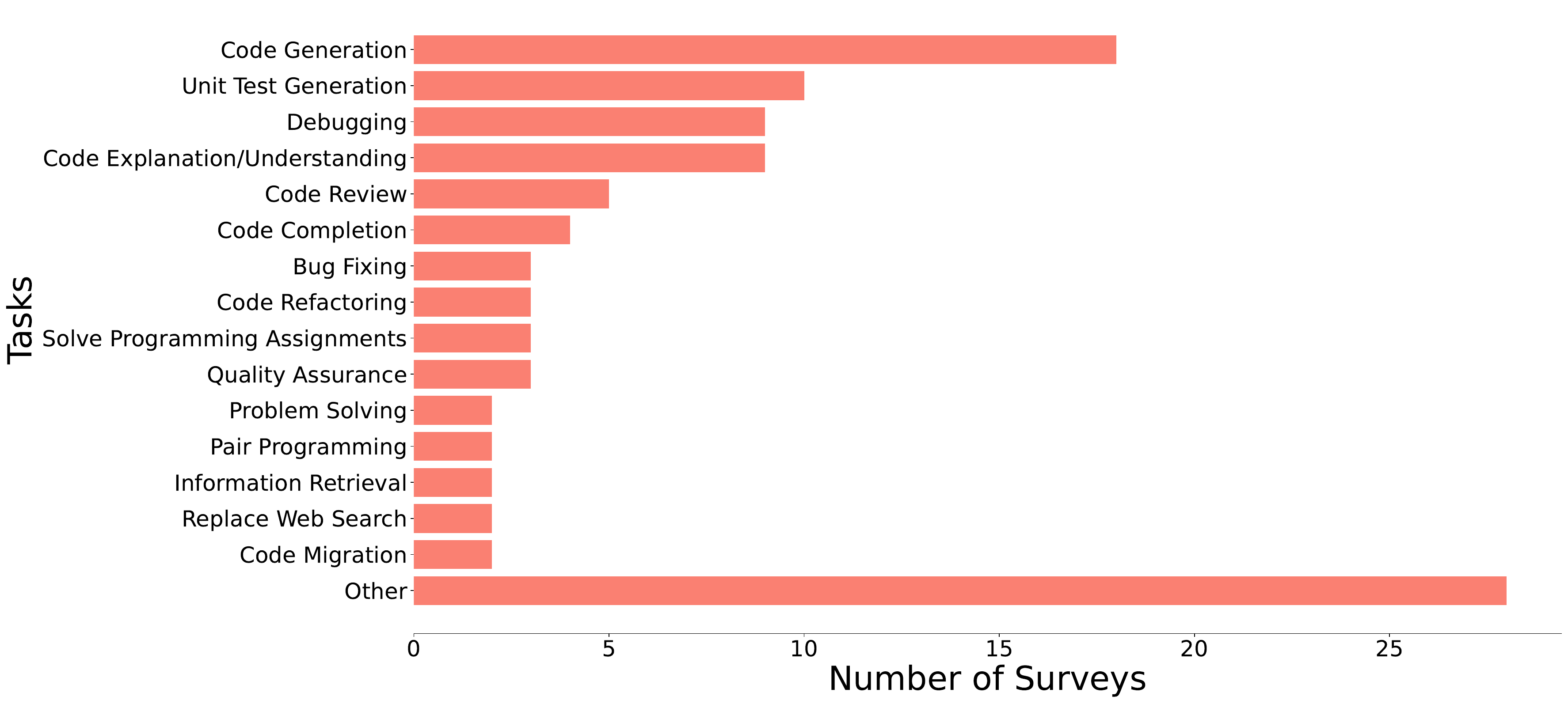}
\caption{Tasks}
\label{fig:comparison_tasks_distribution}
\end{subfigure}
\begin{subfigure}{0.45\textwidth}
\centering
\includegraphics[width=\linewidth]{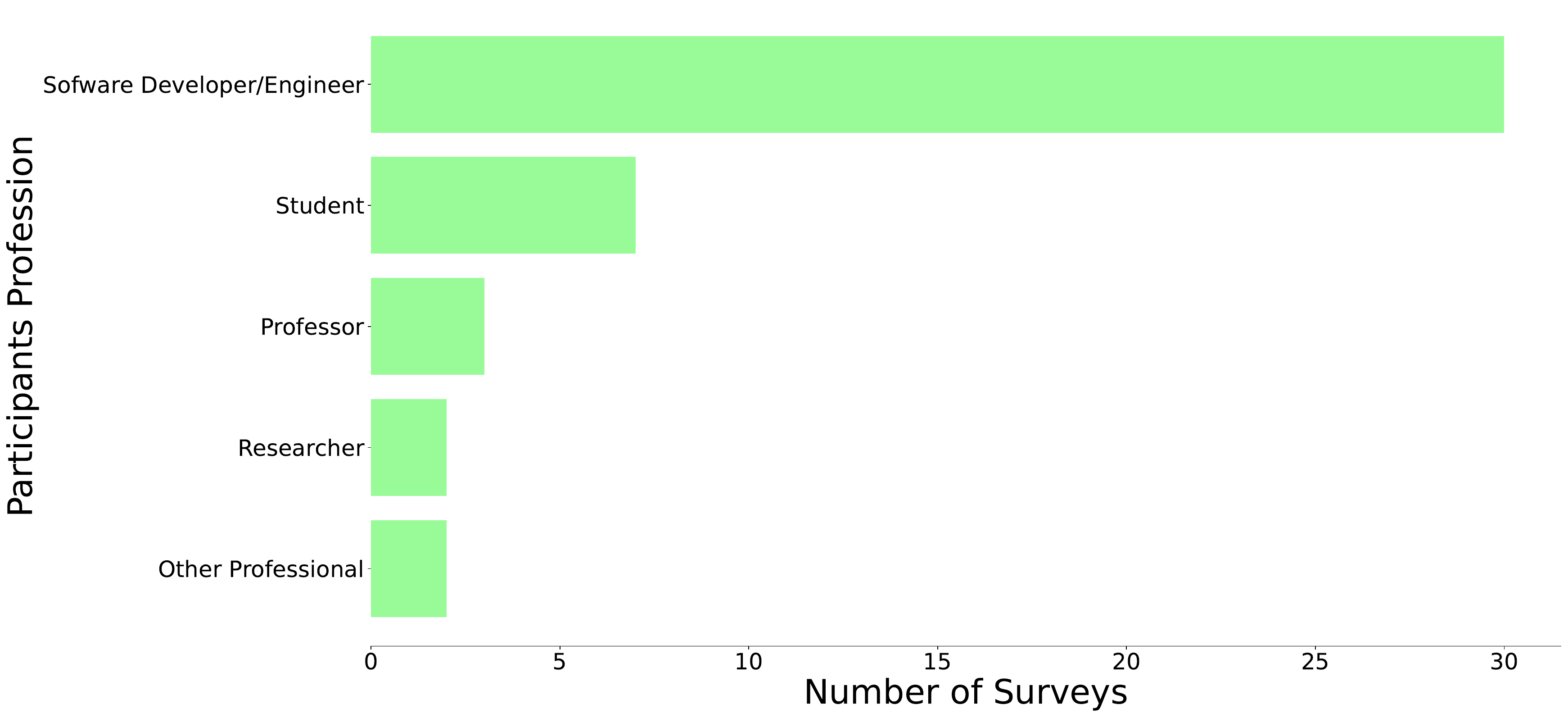}
\caption{User Profession}
\label{fig:comparison_participants_distribution}
\end{subfigure}
\begin{subfigure}{0.45\textwidth}
\centering
\includegraphics[width=\linewidth]{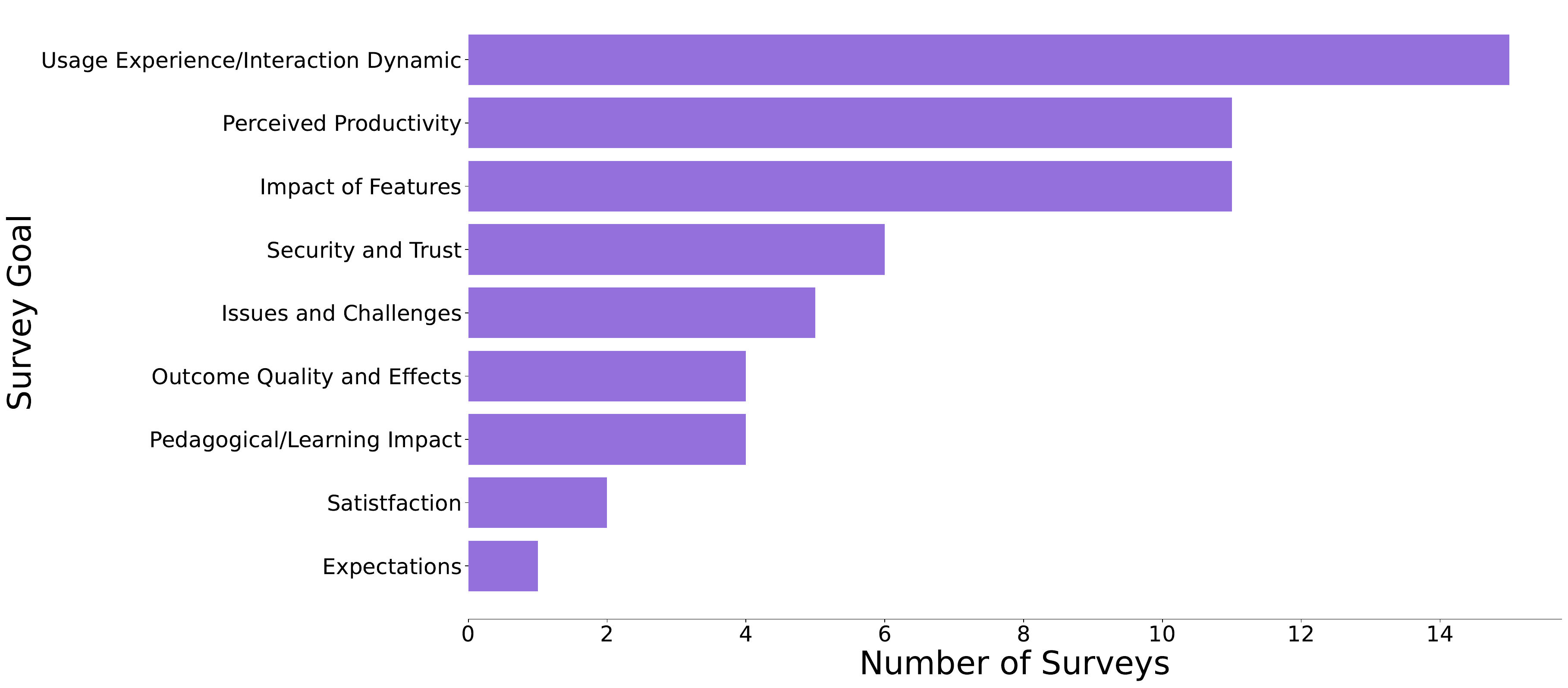}
\caption{Survey Goals}
\label{fig:comparison_goals_distribution}
\end{subfigure}
\caption{Distribution of AI Tools (a), Tasks (b) Users Professions (c) and Goals (d) covered in the 35 SE surveys analyzed.}
\end{figure*}

\subsection{Search Selection Criteria}
We started from a web search using terms such as ``survey'', ``study'', ``user experience'', ``insights'' in correlation with ``AI coding assistant''. From the initial pool of candidate papers, we further expanded the search to any works citing those papers.
The selection criteria for the works to include in this review were then the following.

    \noindent \textbf{Recent}. As the landscape of the features and capabilities of AI coding assistants and the models powering them is progressing at staggering speed, we decided to include only studies from the past two years (2024 and 2025), with two exceptions from 2023 covering the initial opinions of SE using AI coding assistants.
    
    \noindent \textbf{Practitioners Surveys}. We excluded works that contained empirical studies analyzing the features and capabilities of AI coding assistants, but did not include human surveys. Our goal was to understand what has been asked of practitioners using AI coding assistants and which conclusions can be drawn from it.
    
    \noindent \textbf{Specifically Covering SE Tasks}. Our goal is to focus on the perceptions, usage and vision for AI coding assistants specifically for software engineering. Some works were excluded since they surveyed practitioners on other aspects of AI coding assistants.
    
    \noindent \textbf{Publicly Available}. We included works that were accessible through Google Scholar, ArXiv and conferences and journal sites. We excluded enterprise works not publicly disclosed (the only report from a private consulting company \cite{mckinsey2023unleashing} was publicly available).

From an initial pool of approximately 50 papers we retained 35 published between 2023 (2), 2024 (14) and 2025 (19).

\subsection{Methodology}

For each survey, we designed to compare and summarize the following information: which AI Tool(s) it investigated, the number and participants and their background (role, expertise, industry, proficiency levels), the software engineering tasks covered, the overall goal of the study and finally the key findings. 
In order to facilitate the processing of relevant information from the pool of selected papers, we used two frontier models to analyze them at scale, namely Gemini 2.5 pro and Claude Sonnet 4. We prompted each model to act as a research scientist and provide the best prompt for itself to extract the information we designed from each of the papers.

We report the resulting prompts generated by each model in Tables \ref{tab:gemini_prompt} and \ref{tab:claude_prompt} of the Supplementary Material. 

We then proceeded to iteratively prompt each model for each of the papers selected. The information for each paper was then manually verified and summarized in Tables \ref{tab:research_compilation1}, \ref{tab:research_compilation2} and \ref{tab:research_compilation3} of the Supplementary Material.
We then performed an aggregated analysis on the extracted information,   resulting in the following conclusions.

\subsection{Analysis, Gaps and Improvement Areas} 
From the analysis of the surveys outcomes, productivity gains has been identified as the number one perceived benefit. This stems from being able to automate and/or accelerate tedious repetitive tasks as well as simple modifications to existing code. Rapid prototyping and learning opportunities emerge also as clear perceived benefits of surveyed solutions.

While the large (and growing) quantity of user studies is providing useful insights in usage and perceived utility of AI coding assistants, our analysis of prior studies reveals significant gaps. 
First, the empirical base is dominated by ChatGPT and GitHub Copilot (as clearly evident by the distribution in Figure \ref{fig:comparison_ai_assistants_distribution}), leaving comparative understanding of other assistants limited.

Second, most of the studies we found are short-term, focusing on immediate productivity or single specific tasks rather than long-term maintainability, software quality, organizational outcomes or desired new features. Only one study surveyed developers expectations as opposed to experience and outcomes (as seen in the Goals breakdown in Figure \ref{fig:comparison_goals_distribution}). 
Looking at the distribution of tasks reported in Figure \ref{fig:comparison_tasks_distribution}, we can observe how Code Generation, Unit Test Generation, Debugging and Code Explanation/Understanding dominate. Quality assurance is part of only four surveys. This could be a result of both the features offered/advertised so far by AI coding Assistants as well as the limitations of the surveys in asking desiderata from the participants, as opposed to experience with existing functionalities.
The long tail of ``Other'' tasks mentioned in only one survey includes: Diff Verification, Conditional Logic, Commits Writing, Code Selection, Issue Ideation, Expedite Manual Work, Functionality Verification, Implementing New Features, Next-step Hints Generation, Jump-start First Draft, Issue Synthesis, Note Taking, Accelerate Updates, Advanced Reasoning, Code Readability, Reduced Development Time, Pull Requests, Program Repair, Security-related Programming Tasks, Smart Paste, Tackle New Challenges, Testing, Threat Modeling, Trustworthiness Assessment, Workflow Automation, and General Writing Assistance. 

Third, research participants are drawn from homogeneous cohorts, limiting generalization across industries, regions, and demographics. Though studies showed high variability in participant numbers (10 to 17,420) and experience levels (novice to 15+ years), few selected cohorts across different industries or geographies. Very few studies examined specialized groups beyond software engineers and developers: six focused on students using AI tools for learning coding and solving programming puzzles, and two included researchers, as shown in Figure \ref{fig:comparison_participants_distribution}.  None compared the use and desired features of AI coding assistants across groups even within the same company, but working on different divisions and roles. We intend to fill this gap with Research Question 1 in our own survey in Section \ref{sec:rq}.

Fourth, there is a notable absence of analysis targeted to agentic workflows. Given the recent emergence of agentic-first coding assistants offered as specific IDEs or plugins (Cursor, RooCode, Windsurf, Replit, etc.) and CLI options (Gemini cli, Claude cli, OpenAI codex, etc.) the lack of analysis of users experience and expectations specifically around agentic capabilities is a gap that needs to be filled. We call on the SE Research community to fill such gap.

Finally, while trust, governance, and ethical risks are widely acknowledged in the literature \cite{gu2025tasks, jackson2024impact, terragni2025future}, systematic evaluation of solutions—such as personalization, explainability, or policy frameworks remains scarce. These omissions highlight the need for more diverse, longitudinal, and governance-focused studies to inform both research and practice.


\section{Survey}
\label{sec:results}

\begin{figure*}[t]
\centering
\begin{subfigure}{0.3\textwidth}
    \centering
    \includegraphics[width=0.61\linewidth]{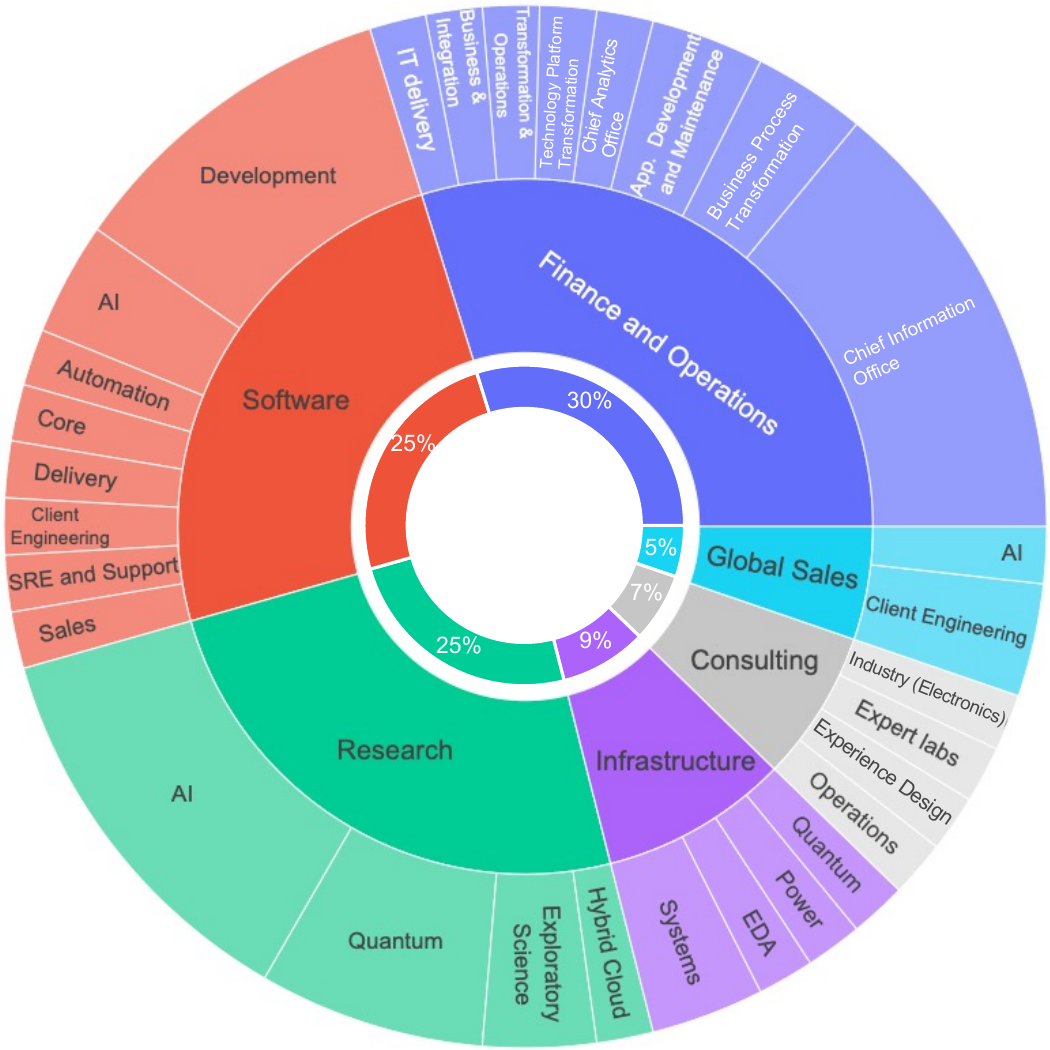}
    \caption{Breakdown of users business units and divisions.}
    \label{fig:users_divisions}
\end{subfigure}
\hfill
\begin{subfigure}{0.3\textwidth}
    \centering
    \includegraphics[width=\linewidth]{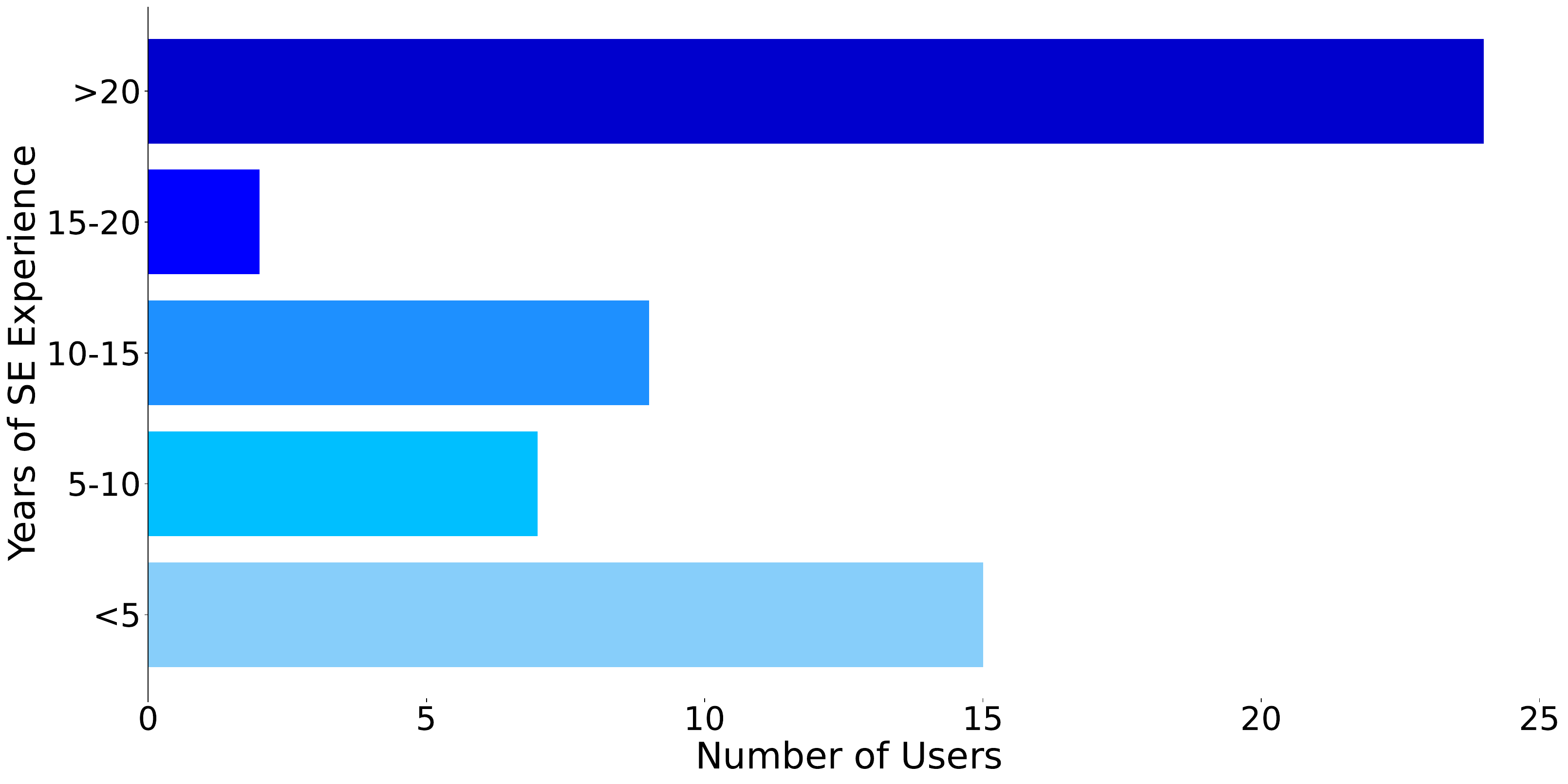}
    \caption{Breakdown of users years of experience in software engineering.}
    \label{fig:years_experience}
\end{subfigure}
\hfill
\begin{subfigure}{0.3\textwidth}
    \centering
    \includegraphics[width=\linewidth]{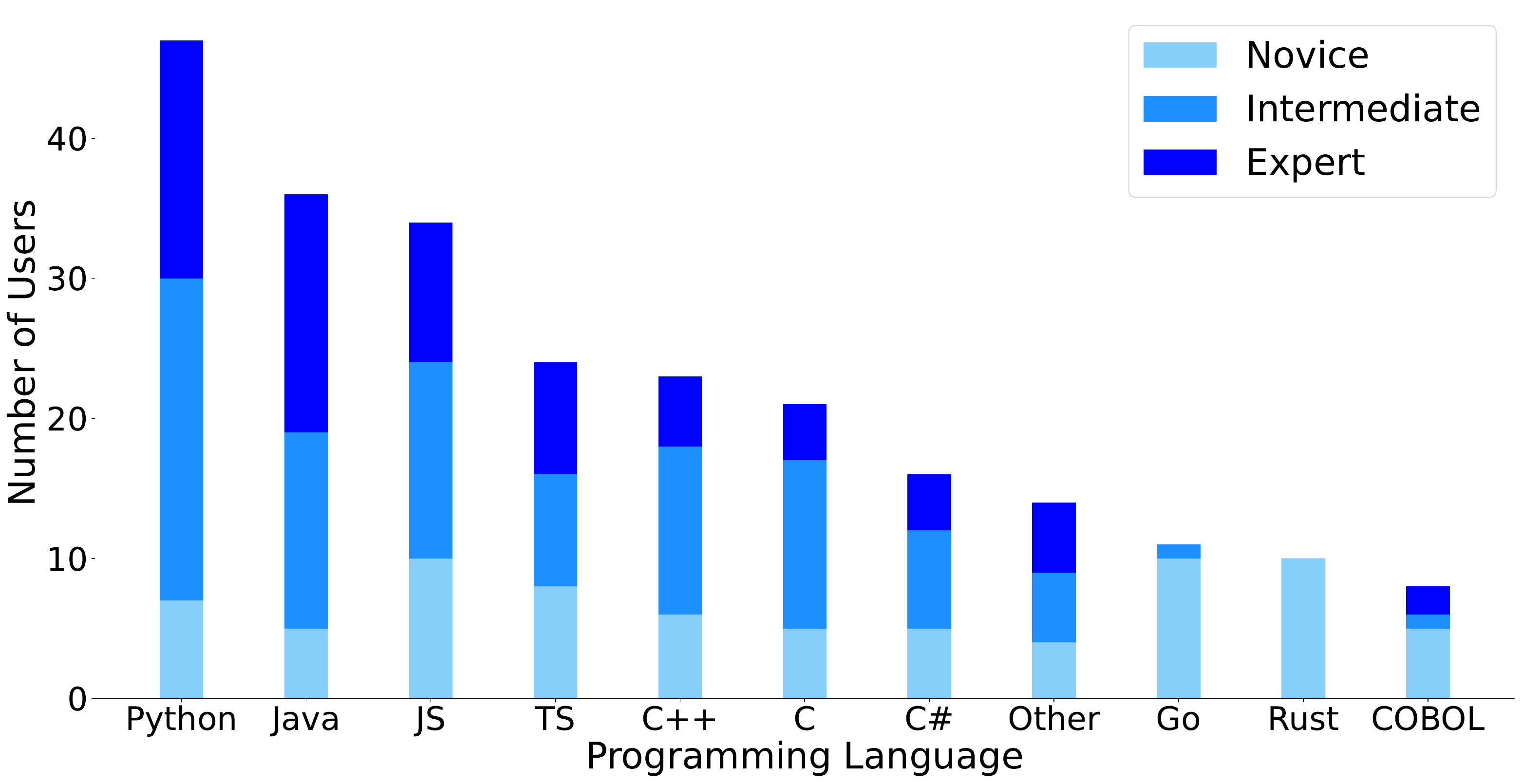}
    \caption{Breakdown of programming languages employed by the users and level of proficiency in each.}
    \label{fig:programming_languages}
\end{subfigure}
\caption{Overview of user demographics and technical background.}
\label{fig:all_users}
\end{figure*}

\subsection{Questionnaire}
We developed a comprehensive survey, with 25 questions, capturing a wide range of data regarding not only the user experience with AI coding assistants, but also practitioners expectations on existing and potentially new features that such assistants can provide, specifically tailored to their workstreams. We launched  data collection in May of 2025 within a tech company, and received 57 responses. Data was collected through an on-line form.

\subsection{Participant Demographics}
In this section, we discuss the demographics of the participants. In total, we received 57 responses from a significantly diverse population. Figures~\ref{fig:users_divisions}, \ref{fig:years_experience}, and \ref{fig:programming_languages} illustrate the key characteristics of the participant pool. Overall, the participants represent a range of departments, including Finance and Operations (30\%), Software (25\%), and Research (25\%), among others. We also found that the participants are highly experienced, with 42\% having more than 20 years of experience. At the same time, the pool includes new hires, with 26\% reporting less than 5 years of experience. Regarding programming language background, while most participants reported using AI assistants for Python, Java, and JavaScript, a significant portion also reported using coding assistants for other languages such as C, Go, Rust, COBOL, C++, C\#, and others.

\begin{figure}[ht]
\centering
\includegraphics[width=0.9\linewidth]{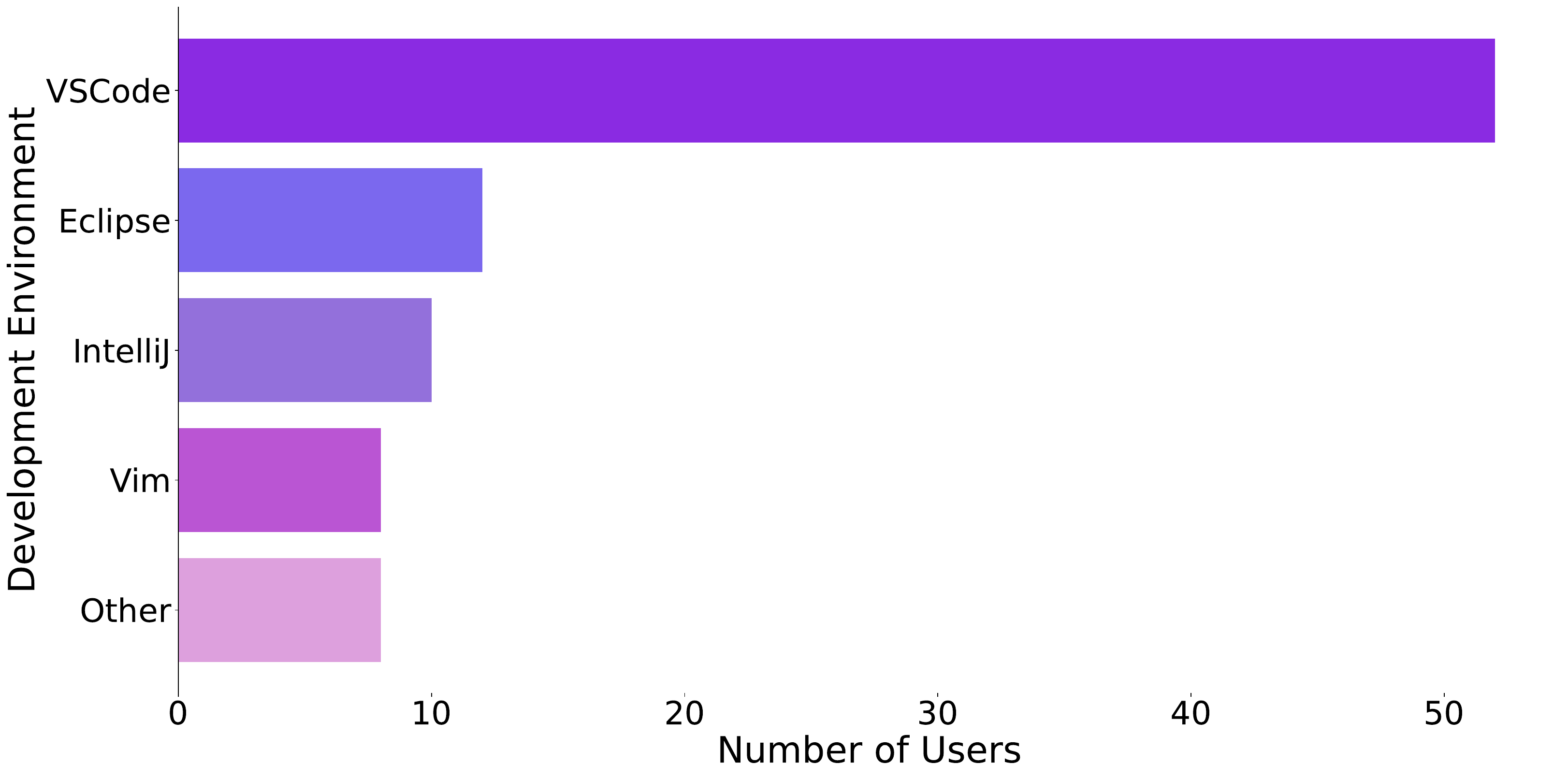}
\caption{Breakdown programming environments and IDEs the user utilize in their daily development (users were allowed to list up to three).}
\label{fig:ides}
\end{figure}

\begin{figure}[ht]
\centering
\includegraphics[width=0.9\linewidth]{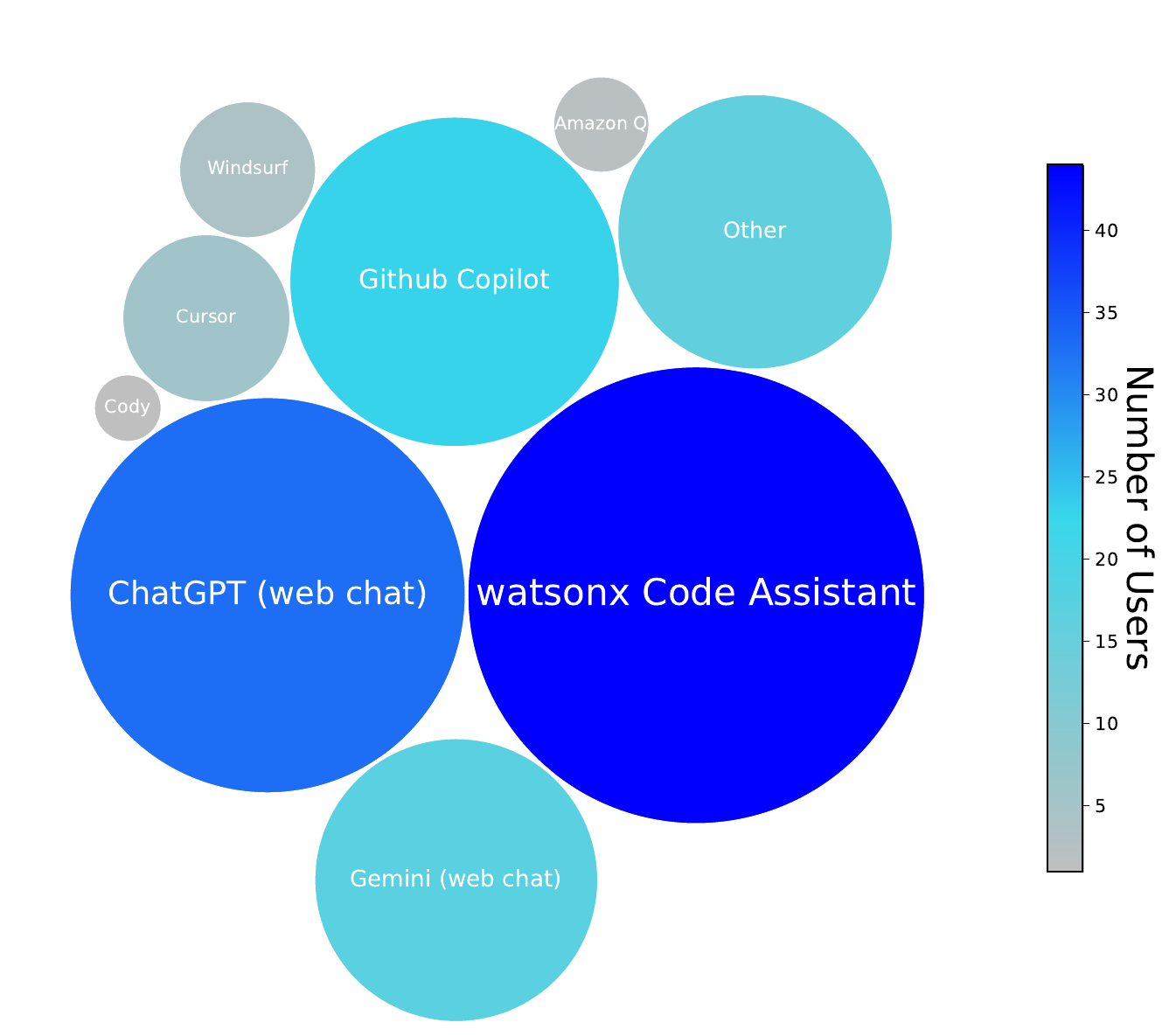}
\caption{Distribution of AI coding assistants with which the users have had hands on experience in the past (users were allowed to list up to three).}
\label{fig:ai_assistants}
\end{figure}

\subsection{Research Questions}
\label{sec:rq}








We designed several questions to assess not only the overall perceived effectiveness of AI coding assistants, but also desired features and unique uses across divisions.

\begin{enumerate}[leftmargin=*]
\item What was the motivation for you to use the AI coding assistant?
\item Which type of help do you benefit the most from it?
\item What fraction of the AI generated code do you actually keep/use? 
\item Is the use of AI coding assistant a productivity gain for you?
\item How did your software engineering process/development change when you started using an AI coding assistant?
\item What are the features of a future ideal AI coding assistant?
\end{enumerate}



\subsubsection{Motivation to use AI coding assistant}

Users reported various motivations to use an AI based coding assistant.  While some were driven by curiosity about the new technology, most had expectations that it can improve their productivity and shorten development time.  These included work in several phases of development, such as exploring ideas for some new program design, fast prototyping, or speeding up an ongoing project, especially when there is a need for much boilerplate code, common helper functions, or highly repetitive editing and transformation. Others hoped to get help in understanding other people's code, or generating documentation for their existing code.  Some expected to save effort in generating tests, improving their code, refactoring, and finding solutions to issues.  Interestingly, several users mentioned that they wanted advices on best practices in coding, or leveraging the assistant to learn about a new package or an unfamiliar programming language.

\subsubsection{Differences across divisions in motivation, use and desired features}

\begin{figure}[ht]
\centering
\includegraphics[width=\linewidth]{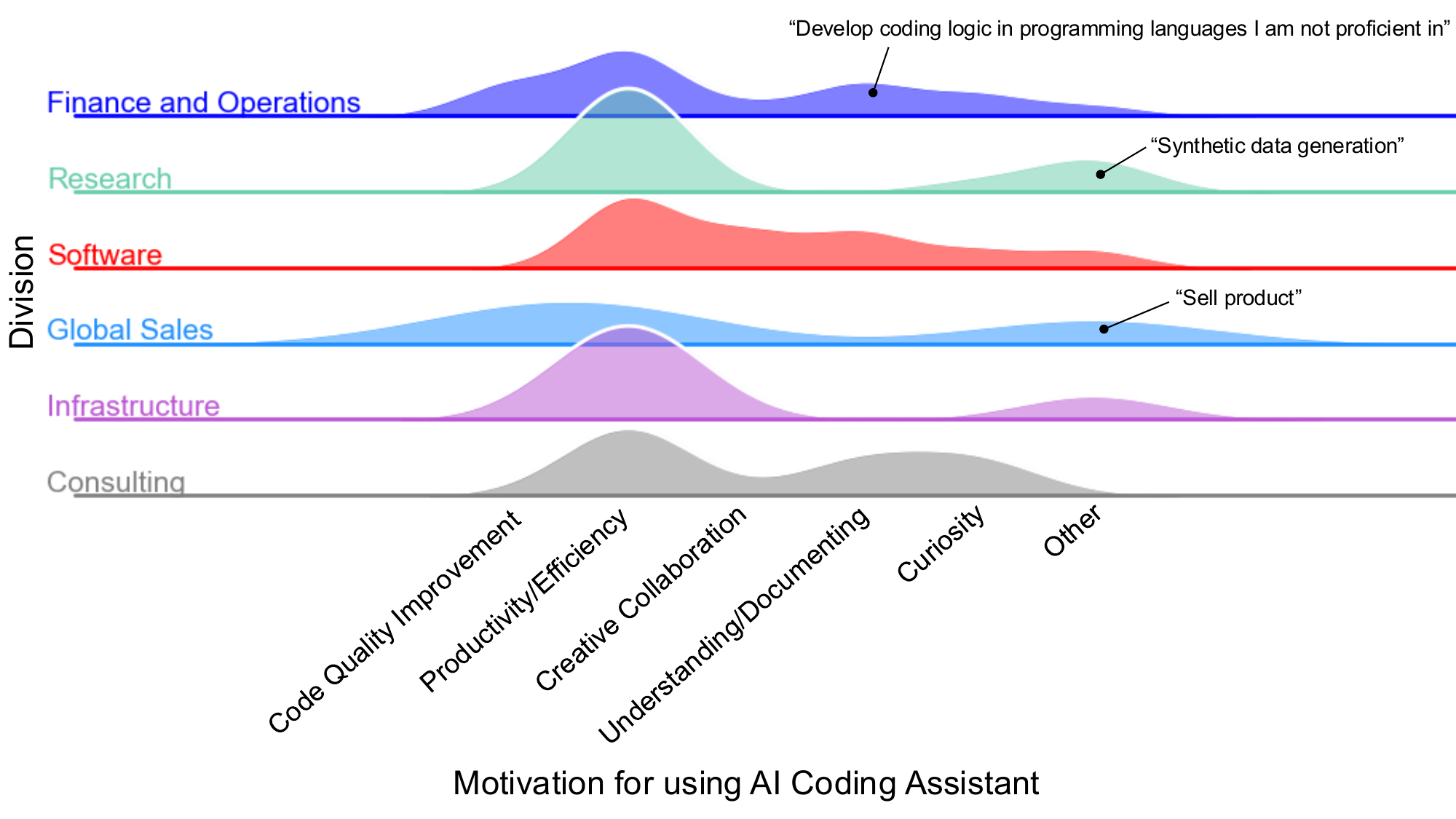}
\caption{Distribution of reasons for using AI coding assistants across divisions.}
\label{fig:motivation_by_division}
\end{figure}

Users from different divisions cited multiple motivations for using an AI coding assistant. We manually mapped the free-text motivations written by users to a set of six general themes: Code Quality Improvement, Productivity/Efficiency, Creative Collaboration, Understanding/Documenting, Curiosity and Other (which includes all answers that could not be mapped to any of the previous themes). While the majority of users in every division employs AI coding assistants for efficiency and productivity gains, from the data reported in Figure \ref{fig:motivation_by_division} we can observe substantial differences across divisions in the distribution of motivations. Within the Research division, productivity is basically the sole motivating use case, while in Global Sales it holds the same importance as improving code quality. Within the Consulting division the understanding code is also a very prominent motivation, whereas in Software we observe a multitude of themes being almost equally cited.
It also is interesting to see very specific examples that were reported in the Other category: for example ``synthetic data generation'' in Research or help to ``sell a product'' in Global Sales. 

\begin{figure}[ht]
\centering
\includegraphics[width=\linewidth]{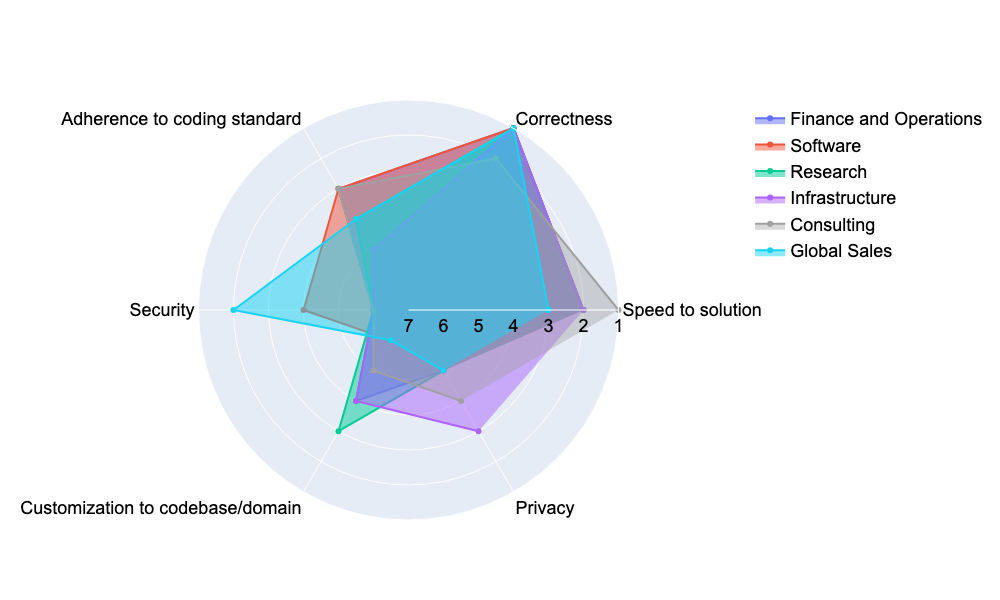}
\caption{Ranking of perceived value of features offered by AI coding assistants, presented by division. Rank of 1 signifies the most valuable feature. }
\label{fig:value_ranking_by_division}
\end{figure}

In terms of perceived value provided by the different features offered by AI coding assistants, we observe large variations in rankings across divisions, as shown in Figure \ref{fig:value_ranking_by_division}. While everyone values \textit{correctness} above all, \textit{security} is the most valuable contribution for Sales, while Research values more the ability to \textit{customize} to very specific codebases and domains, Quantum being a domain mentioned multiple times.

Given the large variety of motivating use cases and adoption types, tailor modes within AI coding assistant are desirable, optimized and targeted for specific use cases.

\begin{tcolorbox}[colback=blue!5!white,colframe=blue!70,boxrule=1pt, parbox=false, title=Motivating Usecases Takeaway]
\begin{enumerate}[leftmargin=0.2in]
\item  There is no ``one size fits all'' motivating use case and adoption type across users of AI coding assistants.
\item Tailor modes/models within AI coding assistants are desirable, optimized and targeted for specific use cases.
\end{enumerate}
\end{tcolorbox}

\subsubsection{Productivity Gains and Code Retention}

\begin{figure}[ht]
    \centering
    \includegraphics[width=\linewidth]{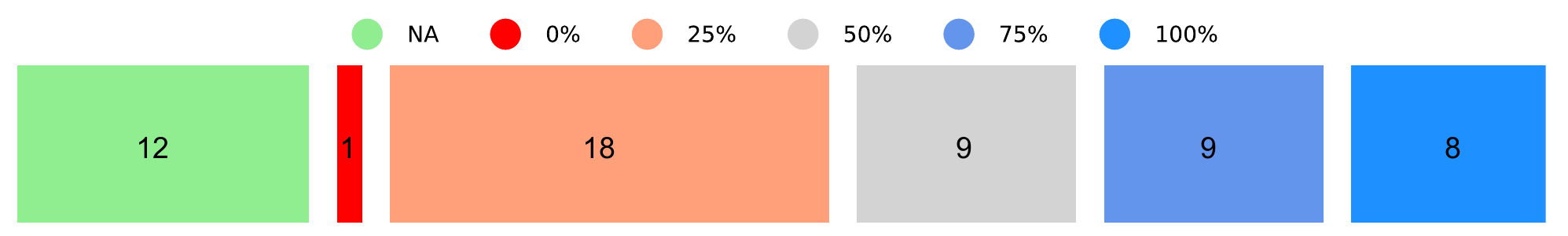}
    \caption{Distribution of Estimated Productivity. The colors represent perceived productivity gain from 0\% (red) to 100\% (blue). 44 out of 57 users (97.8\% of the users who answered on the perceived gain) reported a gain of at least 25\%.}
    \label{fig:productivity}
\end{figure}

The vast majority of respondents found adapting to the use of AI coding assistants easy (39\% somewhat easy, 47\% extremely easy). 88\% of them perceived productivity gains when using AI Coding assistant. Among the 57 responding users, 44 of them reported at least a 25\% increase in productivity.  A smaller fraction (26 users) reported at least a 50\% increase in productivity. 18\% estimate up to 100\% process improvement. We then went on to understand what, if any traditional, non-AI technologies, users have replaced by AI coding assistant.  Over 65\% respondents reported that they previously relied on Google Search, followed by StackOverflow for bug fixes (whether code, config, etc.), and non\-AI IDE completion before transitioning to an AI coding assistant. We then went on to understand user's daily tasks and requirements of a AI coding assistant specific to their role. Out of 35 respondents to this question, 47\% reported that their focus is on code generation (including guidance for coding, code reviews), followed by code understanding and testing.

The majority of users (48) kept the code generated by the AI coding assistant. A large portion of users retained between 25-50\% of the code generated. Many users reporting productivity gains used the AI coding assistant daily or weekly. 

\begin{tcolorbox}[colback=blue!5!white,colframe=blue!70,boxrule=1pt, parbox=false, title=Benefit and Tool Replacement Takeaway]
 Majority of users experience productivity gains from AI coding assistants, primarily in code generation and code understanding; replacing their prior usage of Web tools such as StackOverflow or Google search.
\end{tcolorbox}

\subsubsection{Help received from the AI coding assistant}

Users reporting productivity gains found benefits in both learning and implementation processes.  On the learning side, AI provided knowledge about available packages or coding methods, as well as quick access to available documentation and relevant information; AI also helped in putting together documentation of their code for others to understand.  On the implementation side, AI helped speeding up the development process, from providing suggestions for coding logic, prototyping and scaffolding solutions, generating code drafts, to finishing the full solution via testing and debugging.  

\subsubsection{Change in software engineering process due to using an AI coding assistant}

According to the responders, using an AI code assistant has significantly improved development efficiency and quality. It has become an essential tool for many, helping to (a) follow coding standards and best practices to write clearer code, (b) save time on repetitive tasks and prototyping, (c) quickly identify issues and errors in debugging, (d) ensure compliance with security standards and (e) better understand existing projects and facilitate maintaining and improving them. 
A majority of the surveyed practitioners, 42 out of 57 (73.7\%), reported that the use of AI code assistants has transformed their development process. Specifically it has enabled them to (a) deliver required features and solutions more quickly, (b) focus on high-level design and decision-making, (c) reduce work on repetitive tasks thus saving time for more complex problems and finally (d) try out new technologies and languages with less effort.
The use of AI code assistants varies depending on the context, with some developers using them extensively in personal projects while others are restricted by company policies. 

\subsubsection{Reasons for those users perceiving no gain from using an AI coding assistant}

According to the smaller fraction (13 of 57 (22\%)) of responding users who reported little or no gain in productivity resulting from the use of an AI coding assistant, the reasons are:

\begin{itemize}[leftmargin=*]
    \item Limitations of the coding assistant:  AI can generate boilerplate code but not those that need custom or complex logic.
    \item Correctness and completeness: the code generated by AI may contain errors, omit important features, or not follow standard practices in the codebase.
    \item Prompting needs: a detailed prompt is needed to fully specify the requirements and obtain accurate and relevant code, which is not easier than writing the code manually instead.
\end{itemize}

As a result, they saw negative impact on development, i.e., using AI tools has actually made it harder to debug and fix issues.

A few developers used languages in their current projects that are not supported by the AI coding assistant.  Some preferred to use other IDEs not integrated with an AI coding assistant, and were therefore not able to obtain the expected benefits.

\begin{tcolorbox}[colback=blue!5!white,colframe=blue!70,boxrule=1pt, parbox=false, title=Perceived Helpfulness Takeaway]

AI coding assistants are perceived as helpful to
\begin{enumerate}[leftmargin=0.2in]
\item deliver required features and solutions more quickly.
\item focus on high-level design and decision-making.
\item reduce work on repetitive tasks.
\item try out new technologies and languages more easily.
\end{enumerate}

\end{tcolorbox}

\begin{figure}[t]
    \centering
    \includegraphics[width=\linewidth]{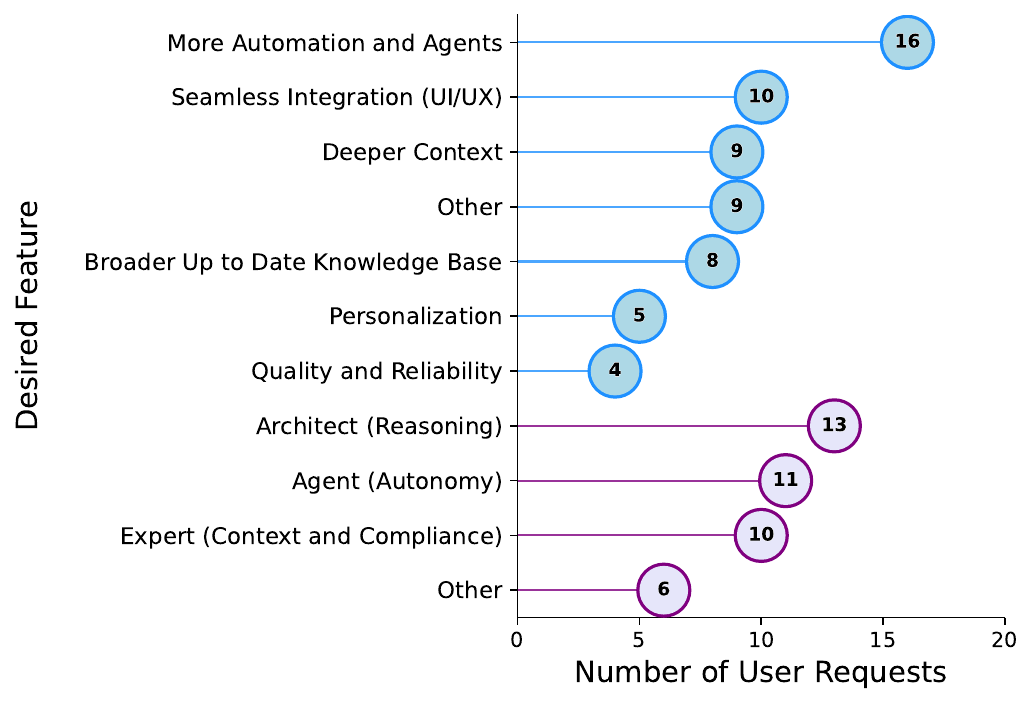}
    \caption{Feature requests for ideal AI coding assistants. Short-term (incremental, top in blue) vs Long-term (transformative, bottom in purple.)}
    \label{fig:short_term}
\end{figure}

\subsubsection{Features of an ideal AI coding assistant}

We asked users to list desirable features not yet offered by current AI coding assistants, distinguishing between short term, that is, perceived as feasible in the near future and more incremental in nature with respect to existing capabilities, and longer term, more transformative directions. The results are reported in Figure \ref{fig:short_term}.

\noindent \textbf{Short-term features}. We mapped the respondents' requests into the following main proposals for incremental improvements. 

\textit{More automation and agents}. The most frequent request is for the assistant to have high quality support for  Git (PRs/merges), CI/CD integration, and issue management, proactive bug detection and automated fixes. Also, interactivity with the user to establish a form of pair programming routine.

\textit{Deeper context}.  Users want the assistant to have deep, full-repository awareness. It must be able to scan and understand their entire local codebase, project files, and documentation without RAG database setup or performance degradation after a certain token count.

\textit{Seamless Integration (UI/UX)}. Users want the assistant to be a part of their core workflow, not a separate tool. This means full integration with their development environment to enable direct, in-line code edits and quick fixes for all tasks.

\textit{Broader and up to date knowledge base}. The assistant's knowledge is seen as too limited. Users demand support for more specific, less popular languages (specifically naming Perl, Shell, Apex, OPL) and up-to-date knowledge of modern SDKs, libraries, and frameworks, including highly functional library version management with monitoring of incompatibilities.

\textit{Quality and reliability}. Improve core quality by being faster and more accurate. A key request is for the assistant to state its confidence level and manifest uncertainty when appropriate, rather than being ``confidently incorrect''.

\textit{Personalization}. Allow users to choose their model (e.g., local vs. remote) and, most importantly, have the assistant learn and adapt to their personal coding style.

Among other suggestions were also to (a) evolve testing capabilities from just unit tests to generating API and UI tests and (b) enable multimodal interaction modes, such as voice based.

\noindent \textbf{Long-term features}.
Shifting the focus from the ``practical present'' to the ``autonomous future'', users were no longer focused on simple plugin/IDE improvements. The three dominant demands are for the AI to evolve from an assistant into an architect, an agent, and an expert. Multimodality the next most required feature. 

\textit{Architect (Reasoning)}. The top request is for an AI that can perform holistic, multi-file architectural design and create complex projects from scratch. High level specifications to full functioning code is also expected, as well as sufficient creative prowess to generate novel algorithms.

\textit{Agent (Autonomy)}. The second-largest theme is a desire for a fully autonomous software engineering agent that can manage the entire life-cycle, from running experiments to making pull requests and monitoring deployments, not only fix issues.

\textit{Expert (Context and Compliance)}. Finally, users want the AI to have deep understanding of two areas.  (a) Full repository understanding. This mirrors a request from short term improvements, underlying how important it is for practitioners. (b) Rules. Specifically, awareness of security, compliance, and governance across geographical regions.

\textbf{Customization} emerged as a strong theme for the ideal AI coding assistant. When asked how important for them is the ability to customize the behavior of AI coding assistant, 50\% responded ``Somewhat important'' and 30\% ``Extremely important''.  In terms of how, coding style, naming conventions, domain-specific knowledge (including best practices / security / compliance ), ever-evolving public libraries have been mentioned.

\begin{tcolorbox}[colback=blue!5!white,colframe=blue!70,boxrule=1pt, parbox=false, title=Future AI Coding Assistant Takeaway]

Future AI coding assistants will require
\begin{enumerate}[leftmargin=0.2in]
\item full customization of models and coding style.
\item optimal context: no size limitation, full repository/project understanding.
\item ability to handle complexity in terms of reasoning, compliance and autonomy in performing tasks for the entire software development life-cycle.
\end{enumerate}

\end{tcolorbox}
\section{Conclusion}
In this paper we presented the results from empirical study of 57 users spanning across multiple enterprise functions (e.g. finance, research, etc.) about their usage and experience with AI coding assistants. Our key takeaways include that the custom CodeLLM models within AI coding assistants are desirable, optimized and targeted for specific use cases. We reaffirmed that the AI coding assistants are helping drive the productivity for the users, irrespective of their programming skill level. 
To complement our study results, we analyzed additional 35 user surveys, to include academic point of view as a comparison to our enterprise-focused study. As AI coding assistants are evolving rapidly, most promise a solution for delivering features more quickly, and reduce work on repetitive tasks. In our future work will tackle the study and understanding of the agentic AI coding solutions and how they change the software development life-cycle.
\bibliographystyle{ACM-Reference-Format}
\bibliography{acmart}
\clearpage

\appendix
\section{Survey}
In this section, we provide a listing of the questions discussed in this survey. 

\subsection{Background}
\begin{enumerate}[leftmargin=0.2in]
    \item What business unit are you part of?
    \item What division are you?
    \item How many years of experience do you have in software engineering? (<5, 5\-10,10\-15,15\-20, >20)
    \item What programming language do you primarily develop in? Please choose maximum 3 with your level of proficiency. For the others, please mark NA. Languages (Python, Java, Javascript, C, Typescript, Go, Rust, C++, C\#, COBOL, Other - Specify); Skill level (Novice, Intermediate, Expert, )NA
    \item What environment or IDE do you use to write code? (VSCode, IntelliJ, Eclipse, Vim, XCode, Other-Specify)
\end{enumerate}

\subsection{Experience with AI Coding Assistants}
\begin{enumerate}[leftmargin=0.2in]
\item Have you used an AI-based coding assistant in the past? (Yes / No)
\item What AI coding assistant have you used in the past? (watsonX Coding Assistant, GitHub CoPilot, Cursor, Windsurf, Cody, ChatGPT (webchat), Gemini (web chat), Amazon Q, AugmentCode (former Tabnine), other-Specify)
\item How often do you use it in your work? (Daily , Weekly, Occaissionaly, Other-Specify)
\item What was the motivation for you to use the AI code assistant? (free text)
\item What type of AI-based coding experience do you prefer? Please rank (IDE based, Editor only, Command Line, WebChat, Other)
\item  Which type of help do you benefit the most from it (select all that apply)?  (Code generation (natural language to code), Test generation,
Code explanation,
Code refactoring,
Code completion,
Chat,
Bug fix,
New feature development Automation of existing code, Knowledge discovery,
Code translation, Application modernization, Other)
\item What fraction of the AI generated code do you actually keep/use? (0\% means you do not keep any code, 100\% you keep all the code) (0, 25,50,75,100)
\item When you think about your usage of AI assisted code assistant, please rank what do you value the most (
Speed to solution,
Correctness,
Adherence to coding standard,
Privacy,
Security,
Customization to own codebase and/or domain Other (please specify in next question))
\item Is the use of AI coding assistant a productivity gain for you? (Y/N)
\item If YES, how much, compared to the existing tools? If NO, please put 0\% below (0,25,50,75,100)
\item If YES, please mention the existing tools or functionalities you referred to (google search, stackoverflow, traditional non-AI IDE code recommendation etc.). If NO, please write NA below (free text)
\item If NO, why not? Please explain.
\item Please rank the type of help from AI code assistant that you have experienced from most valuable to most disappointing, according to your expectations versus experience. (Code generation (natural language to code), Code completion,
Code explanation,
Test generation,
Code refactoring,
New feature development, Chat,
Bug fix,
Code translation, Knowledge discovery)
\item Thinking about your daily usage and requirements of a AI coding assistant, what is specific to your role and or/application? (free text)
\item How did your software engineering process/development change when you started using an AI coding assistant? (free text)
\item What would be the impact of stopping to use of AI coding assistant? (free text)
\item How easy was for you adapting to use an AI coding assistant? (Extremely easy, Somewhat easy, Neutral, Somewhat, hard, Extremely hard)

\end{enumerate}

\subsection{Ideal AI Coding Assistant}
\begin{enumerate}[leftmargin=0.2in]
\item Future AI Code Assistant: Short term incremental functionalities. What would be the specific features that you would like to see in the near future in AI coding assistant, specifically for your role/work-stream? Please focus on features that are not offered already.
\item Future AI Code Assistant: Longer term directions . Given your understanding of the AI technology, which kind of coding help do you believe is beyond reach in near future?
\item If an AI-based coding assistant can be tailored to your needs, in which way do you want it to adapt to them? (for example: coding style, naming conventions, local libraries, domain-specific knowledge)
\item How important for you is the ability to customize the behavior of your AI coding assistant? The customization may include model customization, and customization/adaptation of user experience. (Extremely important, Somewhat important, Neutral, Not important)
\item What kind of input/help will you be willing to give in order to obtain such tailored advantage? (free text)
\item Given the current state of the art of LLM/AI-based code assistants, in what way do you expect to get help from them? (free text)
\item Do you envision using one or more AI coding assistants (and their respective models) depending on your task? (One Coding Assistant with One Model,
One Coding Assistant with Multiple Models,
Multiple Coding Assistants, each with One Model, Multiple Coding Assistants, each with Multiple Models)
\item How do you think coding will evolve in the future, what (if any) new abstractions do you think should be accommodated in the future AI coding assistants? (free text)
\item How do you think coding will evolve in the future, what (if any) new abstractions do you think should be accommodated in the future AI coding assistants? (free text)
\item What new type of interface do you think would be more effective or useful, compared to the existing text based ones? (for example, speech based?) (free text)
\end {enumerate}


\section{Analysis of AI-Powered Coding Tools Surveys}
\label{sec:analysis}



In the following we report the prompts generated by Gemini and Claude to act as Research Scientists extracting relevant information from the 35 survey papers we have analyzed in this study. 

\begin{table}[]
\begin{tcolorbox}[colback=gray!5!white,colframe=gray!70,boxrule=1pt, parbox=false, title=Gemini Prompt]
\textbf{Persona and Task}
You are a senior research scientist specializing in the intersection of AI and software engineering. Your expertise is in evaluating the efficacy and user experience of AI-powered coding tools. Your task is to analyze a given research paper that details a user study on an AI coding assistant and extract specific, structured information from it.

\noindent \textbf{Input}
The full text of a research paper describing a user study on an AI coding assistant.

\noindent \textbf{Output Structure}
Provide a concise, detailed analysis by extracting the following five pieces of information. For each item, provide a clear, bulleted summary.
\begin{enumerate}[leftmargin=0.2in]
\item AI Coding Assistant Studied:
\begin{itemize}[leftmargin=*] 
    \item Identify the specific AI coding assistant (e.g., GitHub Copilot, Amazon CodeWhisperer, a custom tool).
\end{itemize}

\item User Demographics:
\begin{itemize}[leftmargin=*] 
    \item Detail the demographics of the study's participants. Include:
\begin{itemize}[leftmargin=*] 
    \item The total number of participants.
    \item A breakdown of their programming expertise levels (e.g., novice, intermediate, expert, years of experience).
    \item Their primary field of work or study (e.g., software engineering, academia, data science).
\end{itemize}
\end{itemize}

\item Tasks Studied:
\begin{itemize}[leftmargin=*] 
    \item Describe the specific tasks the participants were asked to perform with the AI assistant. State whether the tasks involved:
\begin{itemize}[leftmargin=*] 
    \item Code generation
\item Test generation
\item Code explanation
\item Code summarization
\item Code completion
\item Refactoring
\item Debugging
\item Or other specific tasks mentioned in the paper.
\end{itemize}
\end{itemize}

\item Study Goal:
\begin{itemize}[leftmargin=*] 
    \item State the primary research question or goal of the study. What was the hypothesis the researchers were trying to test? What were they trying to measure (e.g., developer productivity, code quality, user satisfaction, learning curve)?
\end{itemize}

\item Key Takeaways and Conclusions:
\begin{itemize}[leftmargin=*] 
\item Summarize the main findings and conclusions of the study. What were the key results? What did the researchers learn about the use of the AI assistant based on the study's data? What are the main takeaway messages for the broader research community or industry?
\end{itemize}
\end{enumerate}
\end{tcolorbox}
    \caption{Gemini prompt for the analysis of survey paper on AI coding assistants.}
    \label{tab:gemini_prompt}
\end{table}

\begin{table}[]
\begin{tcolorbox}[colback=claudeorange!5!white,colframe=claudeorange!70,boxrule=1pt, parbox=false,left=0.05in,right=0.05in,title=Claude Prompt]
\small
You are a research scientist specializing in AI-based software engineering with expertise in analyzing empirical studies on AI coding assistants. Your task is to systematically extract and synthesize key information from research papers that conduct user studies on AI coding assistants. When provided with a research paper, extract and organize the following information:

\begin{enumerate}[leftmargin=0.15in]
\item AI Coding Assistant(s) Studied
\begin{itemize}[leftmargin=*]
\item Name and version of the tool(s) (e.g., GitHub Copilot, ChatGPT, Claude, Tabnine)
\item Type of assistant (autocomplete, chat-based, inline suggest, etc.)
\item Model details if specified (e.g., GPT-4, Codex)
\item Any customizations or configurations used in the study
\end{itemize}

\item User Demographics
\begin{itemize}[leftmargin=*] 
\item \textbf{Sample size}: Total number of participants
\item \textbf{Expertise level}: Distribution of skill levels (e.g., beginners, students, professionals, experts)
\item \textbf{Experience metrics}: Years of programming experience, familiarity with AI tools
\item \textbf{Field of work}: Industry professionals, academic researchers, students (specify discipline)
\item \textbf{Programming languages}: Primary languages used by participants
\item \textbf{Other relevant demographics}: Age ranges, educational background, geographic location (if reported)
\end{itemize}

\item Tasks Studied
\begin{itemize}[leftmargin=*]
\item \textbf{Task types}: Specify each task category examined (e.g., code generation, debugging, test generation, code explanation, code review, refactoring, documentation)
\item \textbf{Task complexity}: Simple vs. complex tasks, size of code artifacts
\item \textbf{Programming context}: Languages, frameworks, or domains
\item \textbf{Task design}: Real-world vs. synthetic tasks, duration, conditions
\end{itemize}

\item Study Goals and Research Questions
\begin{itemize}[leftmargin=*] 
\item Primary research objectives
\item Specific hypotheses being tested
\item  Metrics and evaluation criteria (e.g., productivity, code quality, correctness, time-to-completion, user satisfaction)
\item  Comparative aspects (e.g., with vs. without AI assistance, comparing multiple tools)
\end{itemize}

\item Takeaways and Conclusions
\begin{itemize}[leftmargin=*] 
\item \textbf{Key findings}: Main empirical results with quantitative outcomes where available
\item \textbf{Performance insights}: How the AI assistant impacted code quality, productivity, or developer experience
\item \textbf{Limitations identified}: Weaknesses or failure modes of the assistant
\item \textbf{User perceptions}: Qualitative feedback, satisfaction, trust issues
\item \textbf{Practical implications}: Recommendations for developers or tool designers
\item \textbf{Future research directions}: Open questions or suggested follow-up studies
\end{itemize}
\end{enumerate}

\noindent \textbf{Output Format}: Present findings in a clear, structured format using the five sections above. When information is not reported in the paper, explicitly state "Not reported" rather than omitting the field. Include relevant quotes or specific metrics when they strengthen the analysis.

\noindent \textbf{Critical Analysis}: Beyond extraction, briefly note any methodological strengths or limitations that affect the validity of the conclusions (e.g., small sample size, limited task diversity, potential biases).

\end{tcolorbox}
    \caption{Claude prompt for the analysis of survey paper on AI coding assistants.}
    \label{tab:claude_prompt}
\end{table}



\clearpage

\begin{sidewaystable}
\vspace*{18cm}
\caption{Comparative Analysis of Research on AI Coding Assistants (Part 1)}
\label{tab:research_compilation1} 
\tiny
\resizebox{\textwidth}{!}{%
 \begin{tabular}{L{1.5cm}L{2cm}L{3cm}L{3cm}L{3cm}L{6cm}}
\toprule
\small{\textbf{Paper}} & \small{\textbf{AI Tool}} & \small{\textbf{Participants}} & \small{\textbf{Tasks}} & \small{\textbf{Goal}} & \small{\textbf{Key Findings}} \\
\bottomrule
Liang et al. (2024) \cite{liang2024large} & GitHub Copilot, Tabnine, ChatGPT, CodeWhisperer, and proprietary tools. & 410 developers (median 6 years experience) from 57 countries. & Survey on experiences across many tasks: Code Generation, Code Completion, Quality Assurance, Debugging, Code Review. & To systematically investigate the prevalence of \textbf{usability factors and challenges} related to AI programming assistants. & \textbullet\ Developers motivated primarily by \textbf{convenience and speed} (e.g., autocompletion 86\%). \newline \textbullet\ Main reasons for non-use: code failed to meet requirements (54\%) and \textbf{difficulty controlling the output} (48\%). \newline \textbullet\ Recommended focusing on better control and chat-based interactions for improved usability. \\
\midrule
Paradis et al. (2025) \cite{paradis2025icseseip} & Google internal IDE and model (Cider). & 96 full-time Google software engineers. & Three complex tasks: AI code completion, smart paste, and natural language to code. & To estimate the impact of three AI features on time spent on a complex, enterprise-grade task. & \textbullet\ Developers who used AI were about \textbf{21\% faster} than those who did not. \newline \textbullet\ Controlling for other factors reduced the statistical significance of the finding. \\ \midrule
Ziegler et al. (2024) \cite{ziegler2024measuring} & GitHub Copilot & 2,631 survey responses; 17,420 users analyzed from Feb. to Mar. 2022. Experience levels ranged from student to 16+ years of professional experience. & Code completion. & To establish a clear link between usage measurements and developer productivity or happiness. & \textbullet\ The \textbf{acceptance rate} is a better predictor of perceived productivity than other measures. \newline \textbullet\ The driving factor for improvements is usefulness as a \textbf{starting point}, not necessarily correctness. \newline \textbullet\ Highly skilled developers may see fewer benefits in writing code with Copilot, but can benefit in other ways. \\ \midrule
Weisz et al. (2025) \cite{weisz2025examininguseimpactai} & IBM watson Coding Assistant & Large-scale survey of wCA users (669); small-scale usability testing (15). & Code explanation, code generation, and unit test generation. & To examine developers' experiences with wCA and its impact on their productivity. & \textbullet\ Productivity gains were unevenly distributed across users. \newline \textbullet\ Understanding code was the top use case, followed by code generation. \\ \midrule
Zhou et al. (2025) \cite{zhou2025exploring} & GitHub Copilot & Empirical study analyzing publicly available data from practitioners on GitHub and Stack Overflow. & General use of Copilot for "AI pair programming" and "code generation." & To empirically understand the \textbf{problems practitioners face} when using GitHub Copilot. & \textbullet\ Most common problems were "Operation Issues" and "Compatibility Issues" stemming from internal errors and network problems. \newline \textbullet\ Developers used generated code to resolve \textbf{5.83\% of issues as-is}, with the rest being modified. \\ \midrule
Sergeyuk et al. (2025) \cite{sergeyuk2025using} & Multiple tools, including ChatGPT, GitHub Copilot, and JetBrains AI Assistant. & 481 professional programmers via a large-scale survey. & Five broad activities: implementing new features, writing tests, bug triaging, refactoring, and writing natural-language artifacts. & To understand how developers use AI assistants, why they don't, what needs to be improved. & \textbullet\ Majority (\textbf{84.2\%}) use at least one AI tool. Most popular are ChatGPT (72.1\%) and Copilot (37.9\%). \newline  \textbullet\ Developers are keen to \textbf{delegate tedious tasks} like writing tests and documentation. \newline \textbullet\ Main reasons for avoiding AI are a lack of trust, company policies, and poor contextual understanding. \\ \midrule
McKinsey Digital (2023) \cite{mckinsey2023unleashing} & Generative AI Tools (Referencing internal studies using tools like GitHub Copilot). & "Software developers," "engineering leaders" (details not provided). & Expediting manual work, Jump-starting first draft, Accelerating updates, Tackling new challenges (e.g., explaining new concepts). & To explore the \textbf{productivity gains} of using generative AI tools and provide a "compass" for technology leaders on deployment. & \textbullet\ Found \textbf{significant productivity gains}: developers complete tasks in about \textbf{half the time} and are 25\%-30\% more likely to complete complex tasks. \newline \textbullet\ Improved \textbf{Developer Experience}: Developers were more than twice as likely to report happiness and a state of flow. \newline \textbullet\ Gains were \textbf{lower for complex tasks} and less experienced developers. \\
\midrule
Bakal et al. (2025) \cite{bakal2025experiencegithubcopilotdeveloper} & GitHub Copilot & A trial with 126 engineers, rolled out to over 400 developers at Zoominfo. & Day-to-day work, including code generation, unit test generation, and code review assistance. & To evaluate the impact of Copilot's deployment on \textbf{developer productivity} and \textbf{satisfaction} in an enterprise setting. & \textbullet\ High developer satisfaction (\textbf{72\%}) and productivity gains (median 20\% time reduction) were found. \newline \textbullet\ Acceptance rate was \textbf{33\%} for suggestions and \textbf{20\%} for lines of code, consistent with other industry reports. \newline \textbullet\ Limitations included a \textbf{lack of domain-specific logic} and inconsistent code quality, emphasizing the need for rigorous code review. \\ 
\midrule
Cui et al. (2024) \cite{cui2024productivity} & GitHub Copilot & 1,974 software developers from Microsoft (1,663) and Accenture (311). & Developers' productivity was measured during their natural, real-world work through: Pull Requests, Successful Builds, Lines of Code Changed, and Commits. & To evaluate the \textbf{productivity impact of Generative AI} in a real-world, enterprise environment. & \textbullet\ Suggestive evidence that Copilot \textbf{increases productivity}, with developers completing more pull requests per week. \newline \textbullet\ The authors note that the estimates are "not very precise" due to experimental challenges. \\ \midrule
Perry et al. (2023)  \cite{Perry_2023} & OpenAI codex-davinci-002 mode & 47 participants with a wide variety of programming experience. & Five security-related programming tasks in Python, JavaScript, and C. & To examine if users write \textbf{more insecure code} when using an AI assistant for security-related tasks. & \textbullet\ Participants with AI access wrote \textbf{significantly less secure code}. \newline \textbullet\ They were more likely to believe their code was secure, indicating \textbf{overconfidence} and a false sense of security. \newline \textbullet\ Users who engaged more with prompts and trusted the AI less produced \textbf{more secure code}. \\ \midrule
Bahn et al. (2025) \cite{bahn2025copiloting} & GenAI tools in general (ChatGPT, GitHub Copilot, etc.). & 18 professionals from 17 European companies (diverse roles: developers, architects, analysts). 1-15 years experience. & Qualitative interview on potential and experienced use cases: Advanced Reasoning, Reduced Development Time, Code Quality Improvement. & To investigate how GenAI can be leveraged in Software Engineering, exploring its \textbf{action potentials and challenges} associated with its adoption. & \textbullet\ Qualitative analysis of expert interviews from 17 European companies. \newline \textbullet\ Identified key use cases: Conceptualization, Code Analysis (error identification/optimization), and replacing search engines. \newline \textbullet\ GenAI offers potential to increase productivity and improve code quality. \newline \textbullet\ \textbf{Challenges}: Reliability, over-reliance ("cognitive misers"), Intellectual Property/data privacy concerns. \newline \textbullet\ Developed a conceptual framework for technology adoption. \\ \midrule
Klemmer et al. (2024) \cite{klemmer2024using} & ChatGPT and GitHub Copilot & 27 software professionals via interviews, plus analysis of Reddit posts. Average experience of 14.6 years. & A wide range of tasks with security implications, from code generation to threat modeling and replacing web search. & To understand the \textbf{human factors} in the use of AI assistants in a security context and what their concerns are. & \textbullet\ A mismatch exists: developers \textbf{mistrust AI's security but rarely find issues}. This suggests they may overestimate their ability to detect flaws. \newline \textbullet\ The main concern for organizations is \textbf{privacy and data leakage}, not the security of the generated code itself. \newline \textbullet\ The developer's role is shifting toward \textbf{supervising AI}, with the human remaining responsible for the final code. \\ \midrule
\end{tabular}}
\end{sidewaystable}
\clearpage

\begin{sidewaystable}[ht]
\caption{Comparative Analysis of Research on AI Coding Assistants (Part 2)}
\label{tab:research_compilation2} 
\tiny
\resizebox{\textwidth}{!}{%
 \begin{tabular}{L{1.5cm}L{2cm}L{2.5cm}L{3cm}L{3cm}L{6.5cm}}
\toprule
\small{\textbf{Paper}} & \small{\textbf{AI Tool}} & \small{\textbf{Participants}} & \small{\textbf{Tasks}} & \small{\textbf{Goal}} & \small{\textbf{Key Findings}} \\
\bottomrule
Takerngsaksiri et al. (2025) \cite{takerngsaksiri2025codereadabilityagelarge} & HULA, an internal Atlassian framework powered by GPT-4 & 118 practitioners via a survey and an empirical case study on internal data. & Survey on perceptions of code readability. Case study comparing code written by humans to code generated by HULA on Jira tasks. & To investigate the role of \textbf{code readability} and compare the readability of AI-generated code to human-written code in a real-world setting. & \textbullet\ A majority of practitioners (\textbf{81\%}) consider readability a critical factor. \newline \textbullet\ HULA-generated code is \textbf{statistically comparable} in readability to human-written code. \\ 
\midrule
Vaz Pereira et al. (2025) \cite{vazpereira2025exploring} & GenAI tools in general, with data on Gemini Code Assist & Software developers from a "large Brazilian media company" (number not specified). & Daily tasks, including writing new code, modifying existing code, and writing test cases. & To understand the real-world experiences, expectations, and challenges of developers using GenAI in a professional context. & \textbullet\ Developers' expectations for improved speed and quality were largely met, with a \textbf{23\% reduction in cycle time}.\newline \textbullet\ The most common use cases were for \textbf{code generation} and \textbf{writing test cases}. \newline \textbullet\ Concerns remain about \textbf{reliability}, \textbf{security}, and the \textbf{impact on collaboration}. \\ \midrule
Martinović et al. (2025) \cite{martinovic2025perceived} & AI-based tools in general, citing GitHub Copilot and ChatGPT. & Software developers from various tech companies (number not specified). & A survey on perceptions of how AI tools affect code quality, productivity, and satisfaction. & To explore the \textbf{perceived impact of AI tools on code quality} from a developer's perspective. & \textbullet\ More than \textbf{75\% of respondents found a positive impact on productivity and satisfaction}. \newline \textbullet\ Perceived impact on code quality (readability, maintainability) was positive but "relatively mediocre." \newline \textbullet\ The study suggests there is significant room for improvement in the quality of AI-generated code. \\ \midrule
Khati et al. (2025) \cite{khati2025mappingtrustterrainllms} & LLMs in general. & 25 domain experts via a survey (students, researchers, engineers, professors). & Surveyed on three tasks: code generation, test case generation, and program repair. & To \textbf{bridge the gap between theoretical and practical understanding of trust} in LLMs for software engineering. & \textbullet\ A significant gap exists between academic definitions of trust and how practitioners perceive it in practice. \newline \textbullet\ Trust is \textbf{highly task-dependent}, with higher trust for test generation (\textbf{75\%}) than for code generation and repair (\textbf{~53\%}). \newline \textbullet\ Developers are primarily responsible for operationalizing trustworthiness. \\ \midrule
Mailach et al. (2025) \cite{mailach2025ok} & Conversational chatbot based on GPT-3.5-turbo & 73 "programming beginners" (students) from a CS2 course. & Solving programming assignments for data structures and algorithms. & To understand how programming beginners use a chatbot and how these \textbf{interaction patterns relate to task performance}. & \textbullet\ Chatbot-assisted students scored \textbf{21\% higher} on average. \newline \textbullet\ Effective patterns involved \textbf{debugging and testing}. Simply generating code was not a successful strategy. \newline \textbullet\ Proper guidance on prompting is \textbf{essential} for beginners to use AI tools effectively. \\ \midrule
Brown et al. (2025) \cite{brown2025howzat} & GPT-4, GPT-3.5, Gemini, Mixtral-8x7B & 44 Java educators (judges) and 5 human experts (benchmark). & Comparative judgment task to rank quality of next-step hints for novice programmers (Java code). & To investigate whether LLMs can generate \textbf{pedagogically useful hints} for novice programmers compared to human experts. & \textbullet\ Educators ranked the quality of LLM-generated hints using comparative judgment. \newline \textbullet\ The study sought to identify which LLM performs best and the most effective prompts for hint generation. \\
\midrule
Prather et al. (2024) \cite{prather2024widening} & GitHub Copilot (IDE integrated) and ChatGPT (conversational chatbot). & 21 students ("novice programmers") from a CS1 course. & C++ programming problem ("More Positive or Negative") involving code generation, conditional logic, and debugging. & To investigate the impact of GenAI tools on novice programmers' problem-solving behavior, focusing on \textbf{metacognition}. & \textbullet\ Analyzed student behavior and thought processes while using AI assistants. \newline \textbullet\ Investigated whether pre-existing metacognitive difficulties persisted or if new difficulties were introduced by GenAI tools. \\
\midrule
Sahni et al. (2025) \cite{sahni2025trainingsocialdynamicsai} & Microsoft 365 (M365) Copilot. & 10 experienced professionals from various industries. & Real-world use cases for M365 Copilot, including writing assistance, notetaking, and information retrieval. & To understand how individual employees \textbf{learn and adopt AI tools} in professional environments, with a focus on informal learning. & \textbullet\ Professionals prefer \textbf{informal learning} (trial-and-error, peer discussions) over formal training. \newline \textbullet\ A significant \textbf{gap exists between perceived efficiency and confidence} in mastering the tool's advanced features. \newline \textbullet\ The tool is most useful for simple, routine tasks like \textbf{writing and summarization}. \\ \midrule
Akhoroz et al. (2025) \cite{akhoroz2025conversationalaicodingassistant} & ChatGPT (90\%), GitHub Copilot (26\%), Claude, Gemini, etc. & 143 student developers (undergrad/grad CS). & Survey on real-world usage: Debugging (73\%), Syntax Explanations (54\%), Problem-Solving (47\%), Code Generation (22\%). & To understand how programmers, specifically students, \textbf{interact with LLM-driven coding assistants} and propose design guidelines. & \textbullet\ Identified common usage patterns and barriers among student developers (C\# and Unity). \newline \textbullet\ Top reported uses were Debugging and Syntax Explanations, not primary Code Generation. \newline \textbullet\ Focused on identifying effective interaction strategies (prompt refinement). \\
\midrule
Wang (2025) \cite{wang2025fromcode} & Copilot for Testing (Context-Based RAG plugin for Xcode). & 12 professional iOS developers. & Subjective evaluation of testing-related tasks: Debugging, Generating test cases, Verifying functionality. & To propose and validate a new software testing methodology powered by a \textbf{context-based RAG module}. & \textbullet\ Introduced a new AI-assisted testing system (Xcode plugin). \newline \textbullet\ Evaluated impact on bug detection accuracy, test coverage, and user acceptance rates. \newline \textbullet\ Compared performance against a baseline model without the RAG mechanism. \\
\midrule
Kuhail et al. (2024) \cite{kuhail2024replace} & Evaluation: ChatGPT 3.5. Survey: Copilot, OpenAI Codex, DeepCode, Polycoder. & 99 professionals (survey sample); ChatGPT 3.5 model for evaluation. & Evaluation: Solving 180 LeetCode problems. Survey: Usage for boilerplate code, explaining code, debugging, test cases. & To assess the \textbf{effectiveness of ChatGPT 3.5} for coding problems and programmers' \textbf{perceptions of capabilities, trust, and job security}. & \textbullet\ Effectiveness: ChatGPT was effective for \textbf{easy and medium} problems, but less reliable for hard problems. \newline \textbullet\ Productivity: 50\% reported increased productivity, correlating with \textbf{higher trust} and heightened \textbf{job security threat perception}. \newline \textbullet\ Limitations: Tools can \textbf{misunderstand requirements} and generate erroneous or inefficient code. \newline \textbullet\ Programmers cite \textbf{domain knowledge} and human ingenuity as key differentiators. \\ \midrule
Stray et al. (2025) \cite{stray2025generative} & GitHub Copilot and ChatGPT. & 14 interviews; observation of two teams (28 members). & Case study in agile environment: Coding, Information Retrieval (replacing search), Problem-Solving, and observing Gen AI use in Solo vs. Pair Programming. & To understand how the integration of Gen AI tools \textbf{influenced a developer's workday} in a professional agile setting (Fintech). & \textbullet\ GAI used more in \textbf{Solo Programming} (improved efficiency/reduced stress); decreased in \textbf{Pair Programming}. \newline \textbullet\ Workflow is now "AI-centric": more reviewing, less writing. \newline \textbullet\ Less experienced developers \textbf{rely more heavily} on AI tools. \\
\midrule
\end{tabular}}
\end{sidewaystable}
\clearpage

\begin{sidewaystable}[ht]
\vspace*{18cm}
\caption{Comparative Analysis of Research on AI Coding Assistants (Part 3)}
\label{tab:research_compilation3} 
\tiny
\resizebox{\textwidth}{!}{%
 \begin{tabular}{L{1.5cm}L{2cm}L{2.5cm}L{3cm}L{3cm}L{6.5cm}}
\toprule
\small{\textbf{Paper}} & \small{\textbf{AI Tool}} & \small{\textbf{Participants}} & \small{\textbf{Tasks}} & \small{\textbf{Goal}} & \small{\textbf{Key Findings}} \\
\bottomrule
Tehrani et al. (2024) \cite{omidvar2024evaluating} & Amazon Q Code Transformation (LLM-based tool for app modernization). & 11 expert Java developers (varied experience: <5 years to >15 years). & Code Migration (Java 8 to Java 17). Reviewing, debugging, and verifying AI-generated diff files in IntelliJ. & To understand the dynamics of \textbf{human-AI partnerships} in complex tasks like code migration and how developers \textbf{build trust} in the outputs. & \textbullet\ Developers act as "directors" and "reviewers," expecting the AI to behave like a \textbf{junior teammate who makes mistakes}. \newline \textbullet\ \textbf{Trust is multi-faceted} (no hallucinations, correctness, completeness). \newline \textbullet\ A \textbf{robust testing suite} is a crucial indicator for establishing trust. \\
\midrule
Nikolov et al. (2025) \cite{nikolov2025googleusingaiinternal} & Google internal LLM & Google Ads PA software engineers (number not specified). & Three specific migrations: JUnit3 to JUnit4, Joda time to Java time, and cleanup of experimental flags. & To share experiences in applying LLM-based code migration and measure if it \textbf{accelerates task completion} by at least 50\%. & \textbullet\ LLMs can \textbf{reduce migration time significantly} but are not sufficient with simple prompting alone. \newline \textbullet\ Success requires a combination of \textbf{AST-based techniques, heuristics, and LLMs}. \newline \textbullet\ The validation and rollout phases are still \textbf{largely human-driven}. \\ 
\midrule
Wang et al. (2024) \cite{wang2024rocks} & ChatGPT (GPT-3.5). & 109 participants (primarily postgraduate students with >1 year professional experience). & Controlled experiment with two tasks: Coding Puzzles and Typical Software Development Task (Bug Fix). & To empirically examine the effects of using ChatGPT in performing typical \textbf{software engineering tasks}. & \textbullet\ Significant \textbf{efficiency improvement} for \textbf{Coding Puzzles}, but only slight improvement for \textbf{Typical Development Task}. \newline \textbullet\ No significant quality increase for either task. \newline \textbullet\ Better outcomes achieved by treating ChatGPT as an active "colleague."\\
\midrule
Das et al. (2024) \cite{das2024developersengagechatgptissuetracker} & ChatGPT (GPT-3.5 and GPT-4). & Developers engaged in 1,152 conversations across 1,012 GitHub issues. & Archival analysis of usage in issue tracker: Ideation (25\%), Synthesis (18\%), Debugging (17\%), Refactoring (13\%). & To investigate how developers use LLMs in a \textbf{real-world, collaborative issue resolution environment}. & \textbullet\ Primary usage is Ideation and Synthesis, but \textbf{reliance on generated code is low} (resolves approx. 12\% of issues). \newline \textbullet\ Developers showed \textbf{dissatisfaction with Debugging} responses (lacking context). \newline \textbullet\ GPT-4 did not show higher reliability or satisfaction than GPT-3.5 in these scenarios. \\
\midrule
Butler et al. (2024) \cite{butler2024deardiaryrandomizedcontrolled} & GitHub Copilot (RCT), also collected data on ChatGPT. & 106 professional software developers (19 junior, 46 senior, 41 principal+). & Diary study observing daily work: Writing boilerplate, Documentation, Replacement for web searches, Learning new codebases, Debugging. & To understand how developers' \textbf{beliefs and self-perceptions change} after regular use in a real-world work environment. & \textbullet\ Increased beliefs that tools are \textbf{useful and enjoyable}, but \textbf{trustworthiness views remained unchanged}. \newline \textbullet\ Developers \textbf{reported higher productivity}, but there were \textbf{no statistically significant changes in objective telemetry metrics} (e.g., LoC, PRs). \newline \textbullet\ Challenges include subtle bugs and need for critical human validation. \\
\midrule
Sabouri et al. (2025) \cite{sabouri2025trust} & LLM-based assistants (GitHub Copilot, PaLM 2, ChatGPT 4). & Survey (n=29: 14 professionals, 15 students); Observation (n=10: 5 professionals, 5 students). Mean experience 4.66 years. & Survey: Rating trustworthiness of code snippets. Observation: Working on self-chosen projects (generating, understanding, selecting code). & To understand how developers make decisions about \textbf{trusting AI-generated code suggestions} (definitions, factors, implications). & \textbullet\ \textbf{High Acceptance (82\%)}, but \textbf{Low Retention (52\%)} in final codebase. \newline \textbullet\ Trust factors cited: \textbf{comprehensibility} and \textbf{correctness}. \newline \textbullet\ \textbf{Automation bias} observed: Trust increased with positive experiences but was not significantly impacted by negative ones. \\
\midrule
Schrijver et al. (2024) \cite{schrijver2024beyond} & JetBrains AI Assistant (GPT-backed) and FLCC (Full Line Code Completion, smaller local LLM). & Metric data from over 26,000 users across 13 JetBrains IDEs. & Analysis of real-world behavior: Micro-level actions, Macro-level actions, and code executions (proxy for testing). & To investigate the impact of LLM-based tools on \textbf{developer behavior} in a real-world setting, focusing on testing and acceptance rates. & \textbullet\ Users with AI tools have \textbf{longer sessions} and \textbf{type more per hour}. \newline \textbullet\ \textbf{Testing behavior decreases}: Code executions (proxy for manual testing) per hour \textbf{decreases} for AI-assisted users (potential risk of over-trust). \newline \textbullet\ FLCC (single-line completion) had a \textbf{higher acceptance rate} than the comprehensive AI Assistant. \\
\midrule
Joe Llerena-Izquierdo et al. (2024) \cite{llerenaizquierdo2024towards} & Tabnine (AI-driven code auto-completion wizard). & 67 students ("Programming Applied to Multimedia"). & Programming learning experience creating web pages (JavaScript, HTML, CSS). Tasks included: Analysis, Implementation, Development. & To contribute to the literature on an experience in \textbf{AI-driven development for programming applied to multimedia}, focusing on learning factors, productivity, and student satisfaction. & \textbullet\ Highly \textbf{positive impact on learning and productivity}, reducing time spent on algorithms and minimizing syntax errors. \newline \textbullet\ High \textbf{student satisfaction} (91\% expressed interest) and improved commitment. \newline \textbullet\ Acknowledged risks of \textbf{over-dependence} and ethical implications in education. \\
\midrule
Stalnaker et al. (2025) \cite{stalnaker2025developer} & GenAI tools in general. & 574 software developers from around the world. & Code generation. & To investigate developers' perceptions of the legal and ethical issues of using GenAI, focusing on \textbf{licensing and copyright}. & \textbullet\ Developers' opinions on legal issues of AI-generated code vary widely. \newline \textbullet\ Many are concerned about \textbf{data leakage} and lack formal legal guidance. \\ \midrule
Huang et al. (2024) \cite{huang2024algorithm} & Amazon CodeWhisperer & Software professionals in a pilot case study (number not specified). & Code generation, automated completion, writing tests, and streamlining workflows. & To explore the impact of generative AI tools on the \textbf{productivity of software professionals} from their own perspective. & \textbullet\ Generally positive perception. Tools are highly valued for streamlining workflows and saving time. \newline \textbullet\ Primary challenges are \textbf{reliability concerns} and difficulty getting desired outcomes, requiring manual fixes. \\ \midrule
Thiago S. Vaillant et al. (2024) \cite{vaillant2024developersperceptionsimpactchatgpt} & ChatGPT and analogous systems (LLMs). & 207 software developers (wide range of experience, 31 countries). & Survey on perceptions across tasks: Coding, Bug fixing, Code review, Debugging, Code explanation, Generating test cases. & To understand developers' \textbf{perceptions of ChatGPT's impact} on their work and on the future \textbf{software job market}. & \textbullet\ 73\% believe ChatGPT \textbf{enhances productivity}. 76\% report positive impact on job satisfaction. \newline \textbullet\ Rated quality \textbf{high for explanation/commenting}, \textbf{low for critical tasks} (bug fixing, security). \newline \textbullet\ Concerns about \textbf{job market impact} (potential job losses vs. new opportunities) and \textbf{raising the barrier to entry} for novices. \\ 
\midrule
\end{tabular}}
\end{sidewaystable}
\clearpage

\end{document}